\documentclass[final,num,12pt]{elsarticle}

\journal{Journal of Theoretical Biology}

\usepackage[utf8]{inputenc}
\usepackage[english]{babel}
\usepackage{amsmath,amssymb,physics,mathtools}
\usepackage{graphicx}% Include figure files
\usepackage{dcolumn}% Align table columns on decimal point
\usepackage{bm}% bold math
\usepackage[mathlines]{lineno}% Enable numbering of text and display math
\usepackage[colorlinks,allcolors=blue]{hyperref}
\usepackage{float}
\usepackage{comment}
\usepackage{soul}

\newcommand{\markasnew}[1]{{#1}}   % don't highlight

\usepackage{xr}
\makeatletter
\newcommand*{\addFileDependency}[1]{% argument=file name and extension
  \typeout{(#1)}
  \@addtofilelist{#1}
  \IfFileExists{#1}{}{\typeout{No file #1.}}
}
\makeatother 

\newcommand*{\myexternaldocument}[1]{
    \externaldocument{#1}
    \addFileDependency{#1.tex}
    \addFileDependency{#1.aux}
}
\myexternaldocument{si}

\begin{document}

\begin{frontmatter}
\title{Artificial selection of communities drives the emergence of structured interactions}

\author[inst1]{Jules Fraboul\corref{corr}}
\cortext[corr]{Corresponding author: jules.fraboul@phys.ens.fr}
\author[inst1]{Giulio Biroli}
\author[inst2,inst3]{Silvia De Monte}

\affiliation[inst1]{organization={Laboratoire de Physique de l’École Normale Supérieure,},
            addressline={ENS, Université PSL, CNRS, Sorbonne Université, Université de Paris}, 
            city={Paris},
            postcode={F-75005}, 
            country={France}}
\affiliation[inst2]{organization={Institut de Biologie de l’ENS (IBENS), Département de Biologie},
            addressline={Ecole normale supérieure, CNRS, INSERM, Université PSL}, 
            city={Paris},
            postcode={75005},
            country={France}}
\affiliation[inst3]{organization={Department of Evolutionary Theory},
            addressline={Max Planck Institute for Evolutionary Biology}, 
            city={Plön},
            country={Germany}}

\begin{abstract}
Species-rich communities, such as the microbiota or microbial ecosystems, provide key functions for human health and climatic resilience. Increasing effort is being dedicated to design experimental protocols for selecting community-level functions of interest. These experiments typically involve selection acting on populations of communities, each of which is composed of multiple species. If numerical simulations started to explore the evolutionary dynamics of this complex, multi-scale system, a comprehensive theoretical understanding of the process of artificial selection of communities is still lacking. Here, we propose a general model for the evolutionary dynamics of communities composed of a large number of interacting species, described by disordered generalised Lotka-Volterra equations. Our analytical and numerical results reveal that selection for scalar community functions leads to the emergence, along an evolutionary trajectory, of a low-dimensional structure in an initially featureless interaction matrix. Such structure reflects the combination of the properties of the ancestral community and of the selective pressure. Our analysis determines how the speed of adaptation scales with the system parameters and the abundance distribution of the evolved communities.
Artificial selection for larger total abundance is thus shown to drive increased levels of mutualism and interaction diversity. 
Inference of the interaction matrix is proposed as a method to  assess the emergence of structured interactions from experimentally accessible measures.
\end{abstract}


\end{frontmatter}
%\linenumbers

\section{Introduction}
Artificial selection has been used for millennia to steer plant and animal characters towards target phenotypes. Recently, it is attracting a lot of interest as a way to control and tune ecosystem services and functions, which are emergent properties of biological communities formed by many different species \citep{hooperEffectsBiodiversityEcosystem2005}.
Particularly interesting in this respect are microbial communities that dispense highly relevant functions, contributing to human health \citep{fujimuraRoleGutMicrobiota2010} as well as to global biogeochemical cycles \citep{katzEvolutionaryTrajectoriesBiogeochemical2004}.
The widespread application of artificial community evolution is nonetheless hampered by the large number of parameters that have potential bearings on the efficiency of the selection protocol, and that must be critically evaluated in designing these experiments \citep{xieSimulationsRevealChallenges2019,arias-sanchezArtificiallySelectingMicrobial2019,changEngineeringComplexCommunities2021a}. Such choices still largely rely on intuition and experience of the experimenter rather than on general design principles. It is therefore difficult to set expectations to be compared with empirical observations. This is particularly important because microbial communities' directed evolution has yielded uneven results \citep{sanchezDirectedEvolutionMicrobial2021}, suggesting that the success of artificial selection may hinge upon some unresolved details of the matching between selection target and ancestral community. 

Numerical simulations of large, virtual communities have started exploring how selection for a collective function affects community composition  \citep{williamsArtificialSelectionSimulated2007,pennModellingArtificialEcosystem2003,arias-sanchezArtificiallySelectingMicrobial2019,changEngineeringComplexCommunities2021a,demonteEcologicalRecipesSelecting2021}. Alternative experimental designs and system parameters have thus be shown to affect the efficiency of the selection process. Given the huge space of possible experimental choices and of interaction types, a fundamental problem is how to asses the robustness of simulation results and use them to optimise selection protocols.

Thorough studies of communities composed of two-species helped identifying key processes involved in artificial selection of communities, and pointed out how competition among composing species may be overcome in attaining collective functions \citep{xieSimulationsRevealChallenges2019,vlietRoleMultilevelSelection2019,xieSteeringEcologicalevolutionaryDynamics2021}. In particular, when community ecology was modelled by two-species competitive Lotka-Volterra equations, evolution of a specific community composition relied essentially on modifications of interspecific interactions \citep{doulcierEcoevolutionaryDynamicsNested2020}. Methods used to attain theoretical insights for communities composed of a few species are however not scalable to more complex communities, where a high number of species coexist. 

Here, we use a general mathematical framework rooted in statistical physics of disordered systems \citep{mezard1987spin} to address the evolutionary dynamics of species-rich communities under artificial selection. Mirroring experimental protocols, communities are selected for an assigned, community-level function of species abundances (e.g. total abundance). Communities that maximise the function get the chance to seed the following generation of communities, that are however 'mutated' with respect to the parental community. The novelty introduced by such mutations fuels open-ended evolution that can reshape the ecology of the evolved communities.
On the ecological time scale (between two selection events - or community generations), we assume species abundances to be described by deterministic equations. 

In the spirit of providing null expectations for species-rich ecosystems with minimal imposed structure \citep{mayWillLargeComplex1972,dobson_going_2020}, we model community ecology by Generalised Lotka-Volterra equations (GLVs) with random interactions. Within this framework, species are characterised by the intensity of intra- and inter-specific pairwise interactions. The statistics of such interactions determine the overall nature of the ecological relationships -- e.g. competitive vs mutualistic -- in the community. The study of disordered GLVs has recently been fuelled by the application of methods from statistical physics \citep{buninEcologicalCommunitiesLotkaVolterra2017,biroliMarginallyStableEquilibria2018,altieri2021properties,garcia2021well}, and has brought important insights into the ecological dynamics of complex communities \citep{barbierGenericAssemblyPatterns2018,hu_emergent_2022}.  
Such models for species-rich communities assume that interactions rates are constant, and focus on the resulting ecological dynamics. 

Here, we address the evolution of the inter-species interaction matrix when selection is imposed on a collective function.
In order to highlight the effect of community-level selection, we chose to represent mutations as a process that randomly changes interactions without biasing the evolution of the target function.
This simplification allows us to analytically derive, in the limit where the ecological and evolutionary time scales are separated, the equation for the matrix dynamics, which captures the effects of community-level selection on interactions. When the function to be maximised is the total abundance, selection drives the emergence of a global mutualistic term akin to collective cross-feeding. Our analytic results predict the interplay of different parameters, including choice of the ancestral community, number of communities and the nature of mutations, in determining the speed and attainability of the target function. This analysis reveals that community-level selection modifies interactions by progressively evolving a complex, structured matrix from an initially featureless one.

\section{Model and Methods}
\markasnew{Before presenting in detail the model, we start by describing the biological problem we want to solve in the manner of a simplified experimental protocol. Let's imagine we want to improve the ability of a microbial community to perform a specific function, such as increasing its biomass or breaking down chemical compounds. We would start by inoculating culture vials with samples of a same initial 'ancestral' community. After an initial growth phase where the abundances of the composing species stabilise, we obtain the first generation of 'adult' communities, that we can then score based on how well they perform the desired function. 

'Newborn' communities of the second generation can be derived from adult communities of the first generation in multiple ways} \citep{changEngineeringComplexCommunities2021a,vessman_novel_2023}.
\markasnew{The simplest community selection method, called 'propagule pool' \mbox{\citep{williamsArtificialSelectionSimulated2007}} consists in choosing the communities with the highest score and letting each of them seed one or multiple newborn communities, without mixing. This ensures that the properties acquired in one generation get inherited by the next generation, except for variations due to mutations, population stochasticity, or sampling at reproduction. 
The same sequence of growth phase, selection and reproduction is repeated over and over again, following the same 'serial transfer' scheme used in artificial selection experiments of microbial populations.
}

\markasnew{Throughout the process, microbes will undergo mutations that can affect the community's ability to perform its function. In particular, these mutations can cause changes in interactions between the different species \mbox{\citep{hansen_evolution_2007,xieSimulationsRevealChallenges2019,doulcierEcoevolutionaryDynamicsNested2020}}, resulting in functional variation between communities, upon which selection can act. Some mutations will be maintained in the communities that survive multiple rounds of artificial selection, and affect in return their ecological dynamics }\citep{pennModellingArtificialEcosystem2003}.

\begin{figure}[h]
\begin{center}
\includegraphics[trim=222 160 20 160,clip, width=11cm]{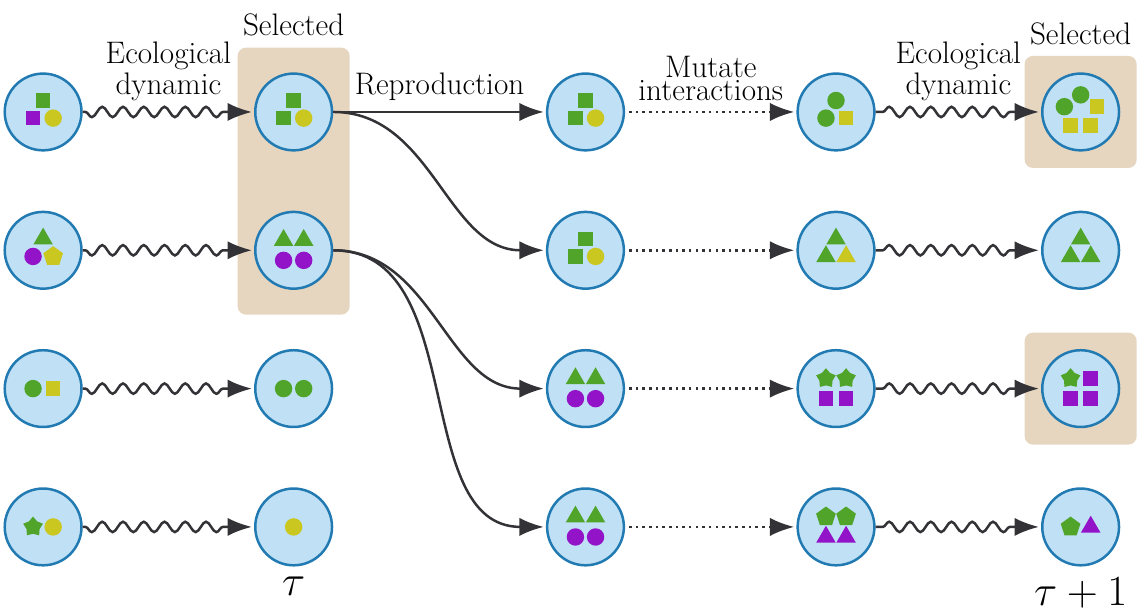} %trim = Left Bottom Right Top
\caption{{\bf Structure of the model for the artificial selection of community function.} Each community in a population of $n$ (here, $n=4$) communities is represented by a circle and is composed of a set of individuals (represented by the \markasnew{tokens}), belonging to different species (represented by colour). The initial, ancestral community is sampled from a same metacommunity, and varies, within a community generation, according to the deterministic ecological dynamics. The $m=2$ communities that at the end of the community generation display the best function (here, largest total population size) are selected for reproduction. Newborn communities of the following generation are generated by copying the state (vector of species abundances), but mutating the parameters of inter-species interactions, as detailed in the text. \markasnew{These changes in interactions, which are different for each newborn community, are represented by the different shapes of the tokens}. This causes community composition to change within the current community generation, at the end of which selection is again applied to the function of the adult community. The same selection-reproduction-ecological growth scheme is repeated at every community generation.
}
\label{fig:evol_dyn}
\end{center}
\end{figure}

\markasnew{Our goal is to describe how community-level selection shapes interactions between species, and how these changes affect the selected function. To this avail, we consider a population of $n$ communities that undergo cycles of ecological growth, selection and reproduction, as illustrated in Fig. {\ref{fig:evol_dyn}}. The ecological dynamics within a cycle is described by a deterministic model, as often done in numerical models that addressed similar questions} \citep{pennModellingArtificialEcosystem2003,williamsArtificialSelectionSimulated2007,changEngineeringComplexCommunities2021a,vessman_novel_2023}. \markasnew{The abundance $N_i$ of any species $i$ belonging to the community is therefore a function of} a continuous time variable $t$. Selection is applied by letting the probability that a community reproduces depend upon a collective function, evaluated at $t=T$, the duration of one community generation.
Reproduction occurs via monoparental seeding of the next community generation ('propagule' reproduction). Community generations are indexed with a discrete variable $\tau$. 
The evolutionary dynamics that we aim to describe consists in the change of the community composition\markasnew{, thus of the species' abundance, across multiple generations. Such changes are associated to the evolution of} ecological parameters, notably inter-specific interactions. For simplicity, we assume that mutations only occur in newborn communities, so that within one collective generation species abundances are only ruled by the ecological dynamics.

In the following, we first detail the model for the dynamics of a single community within one generation, and then the rules for community reproduction and mutation.

\subsection{Within-generation community ecology.}
Each of the $n$ communities is composed of $S$ species with continuous abundances $(N_i)_{i=1,\dots,S}$, whose variation is described by the generalised Lotka-Volterra equations \citep{van_den_berg_ecological_2022}:
\begin{equation}\label{LV_eq}
\frac{\mathrm{d}N_i}{\mathrm{d}t} = \frac{N_i}{K_i}\left(K_i-N_i-\sum_{j\neq i}\alpha_{ij}N_j\right).
\end{equation}
The constants $K_i$ are the carrying capacities and the interaction coefficients $\alpha_{ij}$ represent the effect of species $j$ on the growth of species $i$. 

The carrying capacities, therefore the intra-species interaction strengths, are assumed to be species-specific, and the vector $\vb{K}=\{K_i\}$ does not change over evolutionary time. In the simulations, $K_i$'s are assigned by randomly and independently drawing from a uniform distribution, but the analytical results hold for any vector $\vb{K}$. The matrix of inter-species interactions, on the other hand, is subjected to mutations, starting from an ancestral random matrix, as described below.

\subsection{Species interactions in the ancestral community.}
We choose ancestral communities with random interactions. 
Specifically, the coefficients $\alpha_{ij}$ are drawn from a normal distribution of parameters:
\begin{equation}\label{eq:random}
    \begin{aligned}
    &\mathbb{E}(\alpha_{ij})=\mu/S\\
    &\mathrm{Var} (\alpha_{ij}) = \sigma^2/S \\
    &\mathrm{Corr}(\alpha_{ij}, \alpha_{ji})=\gamma.  \\
    \end{aligned}
\end{equation}

Here, $\mu$ represents the total interaction strength faced by one species from all of its partners, whereas $\sigma$ measures the diversity of interactions.
The parameter $\gamma \in [-1, 1]$ determines the symmetry of the ecological interactions: competition and mutualism correspond to $\gamma=1$ whereas $\gamma=-1$ indicates exploitative interactions like predator-prey and parasitic interactions.

The ecological dynamic of such communities has been characterised in the limit of large number of species $S$ \citep{buninDirectionalityCommunitylevelSelection2021}, where it only depends on the summary statistics of the interaction matrix: $\mu$, $\sigma$ and $\gamma$. Among the three qualitatively different dynamical regimes the system can display, we chose \markasnew{the matrix of the ancestral community} in a region where interactions are competitive and not too diverse, so that the system has a unique, globally stable equilibrium (Supplementary Section \ref{SI:section:phase}).

\subsection[State of the community at the end of a generation]{\markasnew{State of the community at the end of a generation}}
\markasnew{The state of the community at the end of one generation (at time $t=T$) generally depends on the abundances of the newborn community (at time $t=0$). If $T$ is too small for the dynamics to have reached an attractor, the transient composition of adult communities can have, when selection is applied to a community function, unpredictable effects on long-term evolution \mbox{\citep{changEngineeringComplexCommunities2021a}}.
For this reason, we assume that the duration of one generation is large enough for abundances of adult communities to be close to their asymptotic attractor, that is the ecological steady state defined by the interaction matrix at that generation.}

\markasnew{We start from a situation where the ancestral community has a unique, globally stable equilibrium. By its structural stability, small perturbations of the interaction matrix -- as those realised in the first steps of evolution -- will still give rise to stable equilibria. This is however not guaranteed after many generations of community selection, and the stability of the ecological equilibrium may eventually be lost. As we will discuss later, we will focus on the region where the within-generation ecological dynamics has a stable equilibrium.
}

\subsection{Community-level selection and reproduction.}
Selecting communities requires ranking them according to a single collective function. We will essentially focus on the total community abundance $N_T=\sum_i N_i$. Our approach can be generalised to any function $f(\vb{N})$ of the abundances, as we will point out later. The $m$ communities ($m=1$ for the analytical derivation) that at the end of one generation have larger $N_T$ are chosen for reproduction, and the rest is discarded (Fig.\ref{fig:evol_dyn}). Such death and birth processes is what characterises community-level selection. When an offspring community is born, it acquires the same composition of the parent community. In the absence of variation in the community parameters, this guarantees that community functions are perfectly inherited.

\subsection{Community-level mutations.}
For evolution by natural selection to occur at the level of communities, there must be variation in the collective function \citep{lewontinUnitsSelection1970}. 
At each community generation the interactions between species change as some species undergo mutations. Fully characterising the stochastic process associated to these changes is an open challenge. Here, we focus on a simplified model in which these changes, called 'community-level mutations', are random, small and \emph{unbiased}. The latter feature ensures that the collective function does not undergo directional changes unless selection is applied. Although this is a strong simplification, it allows us to study specifically the evolutionary consequences of community-level selection on species interactions. 
In order for mutations not to bias a priori the change of the trait under selection, they need to maintain, \emph{in expectation}, the mean and variance of the interaction \markasnew{matrices of newborn communities}. Even though the expected value of the collective function after mutation remains unchanged, single realisations of the mutation yield however different collective functions, producing the variation \markasnew{between communities} selection acts upon.

We write the interaction matrix at generation $\tau$ as:
\begin{equation}\label{eq:b_mat}
    \alpha_{ij}(\tau)=\frac{\mu(\tau)}{S} + \frac{\sigma(\tau)}{\sqrt{S}} \, b_{ij}
\end{equation}
where: 
\begin{equation*}
\begin{aligned}
\frac{\mu(\tau)}{S} &= \frac{1}{S^2}\sum_{ij}\alpha_{ij}(\tau)\\
\frac{\sigma(\tau)}{\sqrt{S}}& = \sqrt{\frac{1}{S^2}\sum_{ij}\left(\alpha_{ij}(\tau)-\frac{\mu(\tau)}{S}\right)^2} 
\end{aligned}
\end{equation*}
are the \emph{empirical} mean and standard deviation of the matrix $\alpha$, and the reduced matrix $b$ has empirical mean $0$ and empirical variance $1$. 

The mutated interaction matrix \markasnew{of one newborn community} is then defined as:
\begin{equation}
    \alpha_{ij}(\tau+1)=\frac{\mu(\tau)}{S} + \frac{\sigma(\tau)}{\sqrt{S}} \, \hat b_{ij}
\end{equation}
with:
\begin{equation}\label{mut_eq}
    \hat b_{ij} = \frac{b_{ij}+\varepsilon \eta_{ij}}{\sqrt{1+\varepsilon^2}},
\end{equation}
where $\eta$ is \markasnew{a realisation -- different for every $\tau$ and each community -- of} a Gaussian random matrix of expected value $0$, variance $1$ and symmetric correlation $\gamma$. 
Therefore, \markasnew{when averaging over all possible communities of generation $\tau+1$ (thus, over $\eta$), the interaction matrices have the same summary statistics. Mutations therefore don't introduce any bias in between-community variation of interaction matrices, so that interactions get reshaped along an evolutionary trajectory only by the action of community-level selection.}

\subsection{Code description}
Numerical simulations were performed in python using the code accessible at {\small \url{https://github.com/jules-fbl/LV_community_selection}}. 
All the figures of the paper were obtained with a number of species $S=100$, $m=1$ selected communities out of $n=10$, a mutation strength $\varepsilon=0.02$, an initial interaction matrix drawn from a Gaussian distribution of parameters $\mu=3$, $\sigma=0.3$ and $\gamma=0$ and random carrying capacities drawn uniformly between $0.5$ and $1.5$.
The collective generation time was chosen to be  $T=500$ (with the exception of the first generation where a time $T=5000$ was used in order to avoid the propagation of transient effects). This time is long enough for the mutated communities to approach their equilibrium abundances. To integrate the Lotka-Volterra equations, we used an integration scheme described in Supplementary section \ref{SI:numerical_considerations}.
We also imposed an abundance cut-off $N_{\rm{min}}=10^{-20}$ below which species are deemed extinct. \markasnew{This cut-off was added for numerical convenience but has no significant impact on the results.}

\section{Results}

We start by discussing the evolutionary dynamics of the interactions when no selection is applied. 
Then, we present numerical simulations of the model previously introduced to illustrate salient features of the evolutionary dynamics under selection for total abundance. We finally explain the theoretical framework that allows to generalise these observations and outline their scope of application.

\subsection{Community evolution without artificial selection}

As a preamble, it is interesting to study the dynamics induced by community mutations in the absence of selection. As we show in the Supplementary section \ref{SI:random_walk}, the interaction matrix remains Gaussian and is hence completely determined, for large $S$, by its mean and variance. The probability distribution of
$\mu(\tau+1)$ and $\sigma(\tau+1)$,
conditioned on their value at the previous generation ($\mu(\tau)$ and $\sigma(\tau)$), reads, at first order in $\epsilon$:
\begin{equation}
\begin{aligned}
\mu(\tau+1) &\sim \mathcal{N}\left(\mu(\tau), \, \frac{\sigma(\tau)\varepsilon}{\sqrt{S}}\right) \\
\sigma(\tau+1) &\sim \mathcal{N}\left(\sigma(\tau), \, \frac{\sigma(\tau) \varepsilon}{S}\right),
\end{aligned}
\end{equation}
where $\mathcal{N}(\mu,\sigma)$ is a Gaussian distribution of mean $\mu$ and standard deviation $\sigma$.

In the absence of selection, therefore, the summary statistics of the interaction matrix evolve by \emph{neutral drift} and the matrix retains its ancestral random character.
In expectation over different realisation of mutated communities, the interaction matrices at two successive generations have the same summary statistics. As a consequence, any community function will also change by drift. The lack of directionality of evolution in the absence of selection is a consequence of our choice of not representing the biases that can be induced by intra-community selection.

\subsection{Numerical simulations of community evolution under artificial selection}

\begin{figure}[h]
    \centering
    \includegraphics[width=10cm]{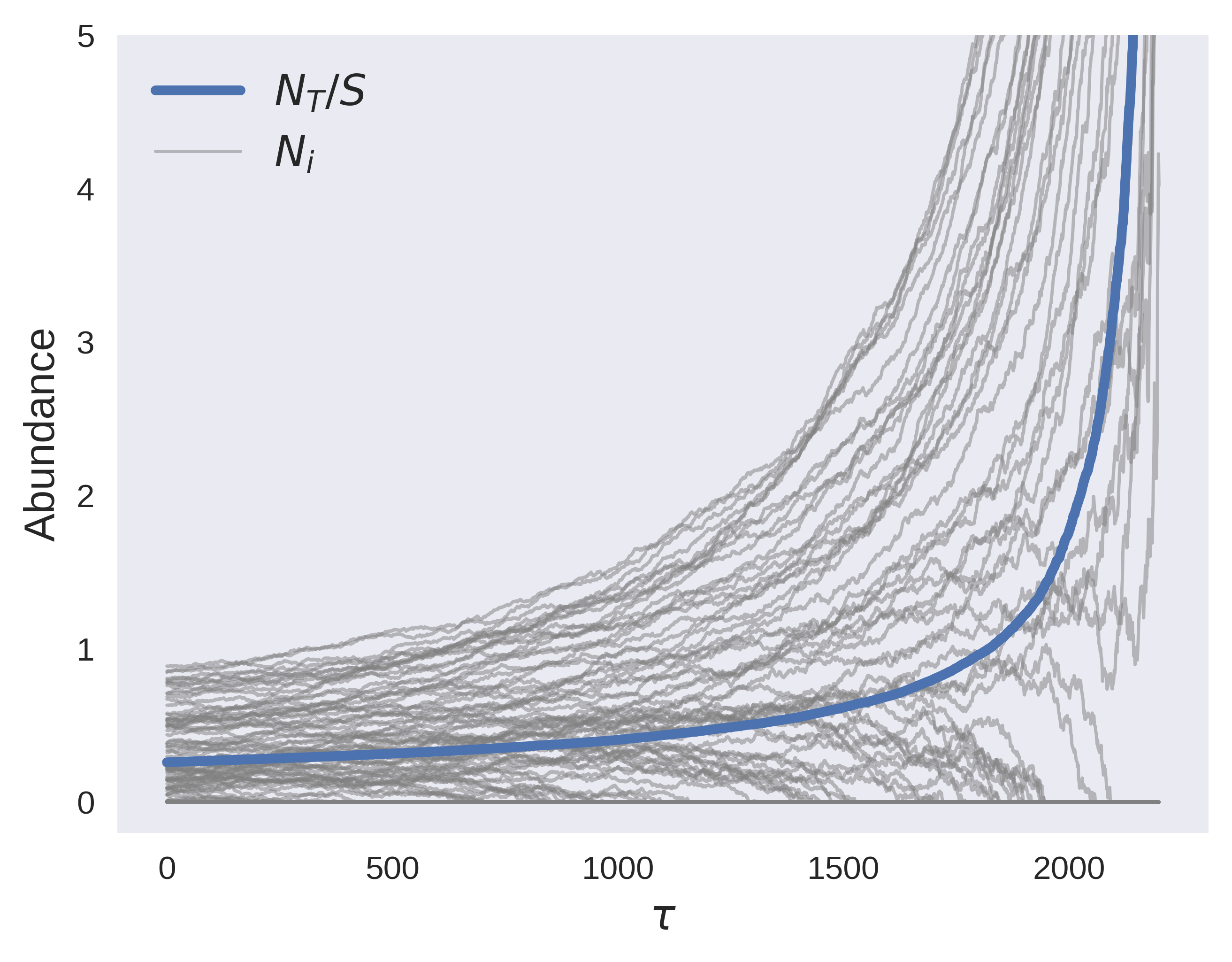}
    \caption{{\bf Changes of species abundance along an evolutionary trajectory.} Selection for increased total abundance leads to an increase in the abundances of most species (grey lines), and, as a consequence, of the average abundance $N_T/S$ (blue line)}
    \label{fig:abundance}
\end{figure}

As observed in past numerical studies  \citep{williamsArtificialSelectionSimulated2007,pennModellingArtificialEcosystem2003,pennRoleNonGeneticChange2004,ikegamiDynamicalSystemsApproach2002}, we find that in response to selection, communities evolve so as to improve the desired collective function (Fig. \ref{fig:abundance}).
In our case, the rate of improvement increases over time, so that the ecological dynamics is eventually pushed in a region where some abundances diverge. Such divergence is a well-known pathology of the Lotka-Volterra equations that can be corrected by choosing a saturation stronger than quadratic \markasnew{\mbox{\citep{sidhom_ecological_2020}}}. We will focus on the regimes where the total abundance increases, but does not diverge. 

\begin{figure*}[h]
    \centering
    \includegraphics[width=16cm]{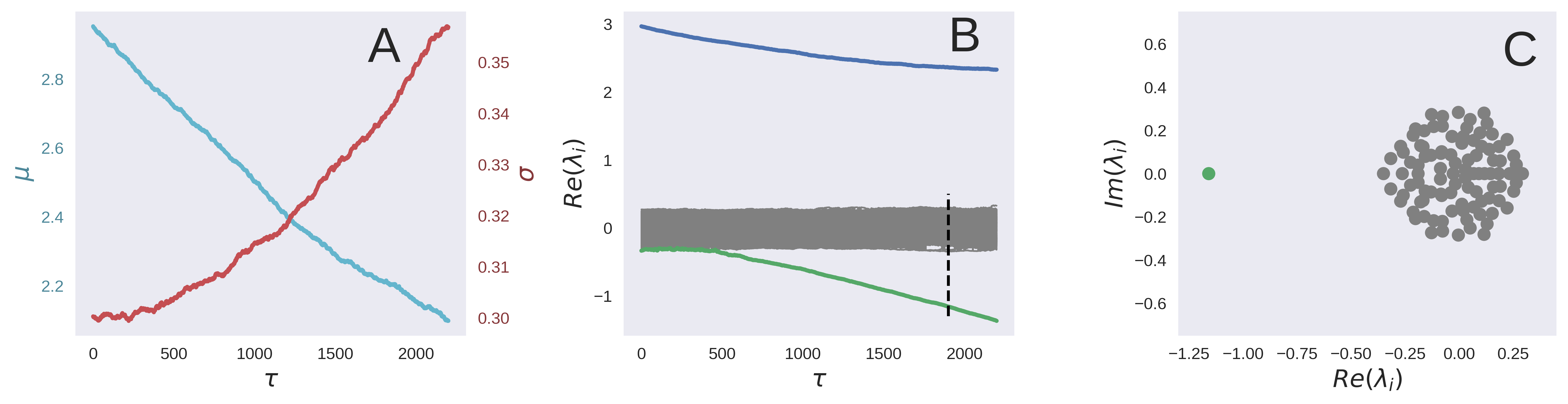}
    \caption{{\bf Changes of the interaction matrix along an evolutionary trajectory.} The interaction matrix $\alpha$ of the best community evolves so that the average interaction strength decreases linearly in time (A, cyan), while its variance increases (A, red). Such changes are accompanied by a modification in the matrix structure, manifest in the spectrum of its eigenvalues. The dynamics of their real part across community generations (B) reveals the appearance of an isolated negative real eigenvalue (green), as well as the decrease of the eigenvalue associated to $\mu$ (blue). A zoom of the spectrum in the complex plane (C) at generation $\tau=1900$ (represented by the dotted line in (B)) reveals that, apart from the emergence of this mutualistic collective mode, the matrix retains its initial random structure characterised by a circular eigenvalues distribution.}
    \label{fig:stats}
\end{figure*}

The observed improvement of community function derives from changes of the interaction matrix $\alpha$, that is also visible on its empirical statistics $\mu(\tau)$ and $\sigma(\tau)$.
As shown in Fig. \ref{fig:stats} A, the mean decreases, indicating that interactions become -- on average -- progressively more mutualistic. At the same time, their variance increases, so that interactions within the community become more diverse. 

\begin{figure}[h]
    \centering
    \includegraphics[width=10cm]{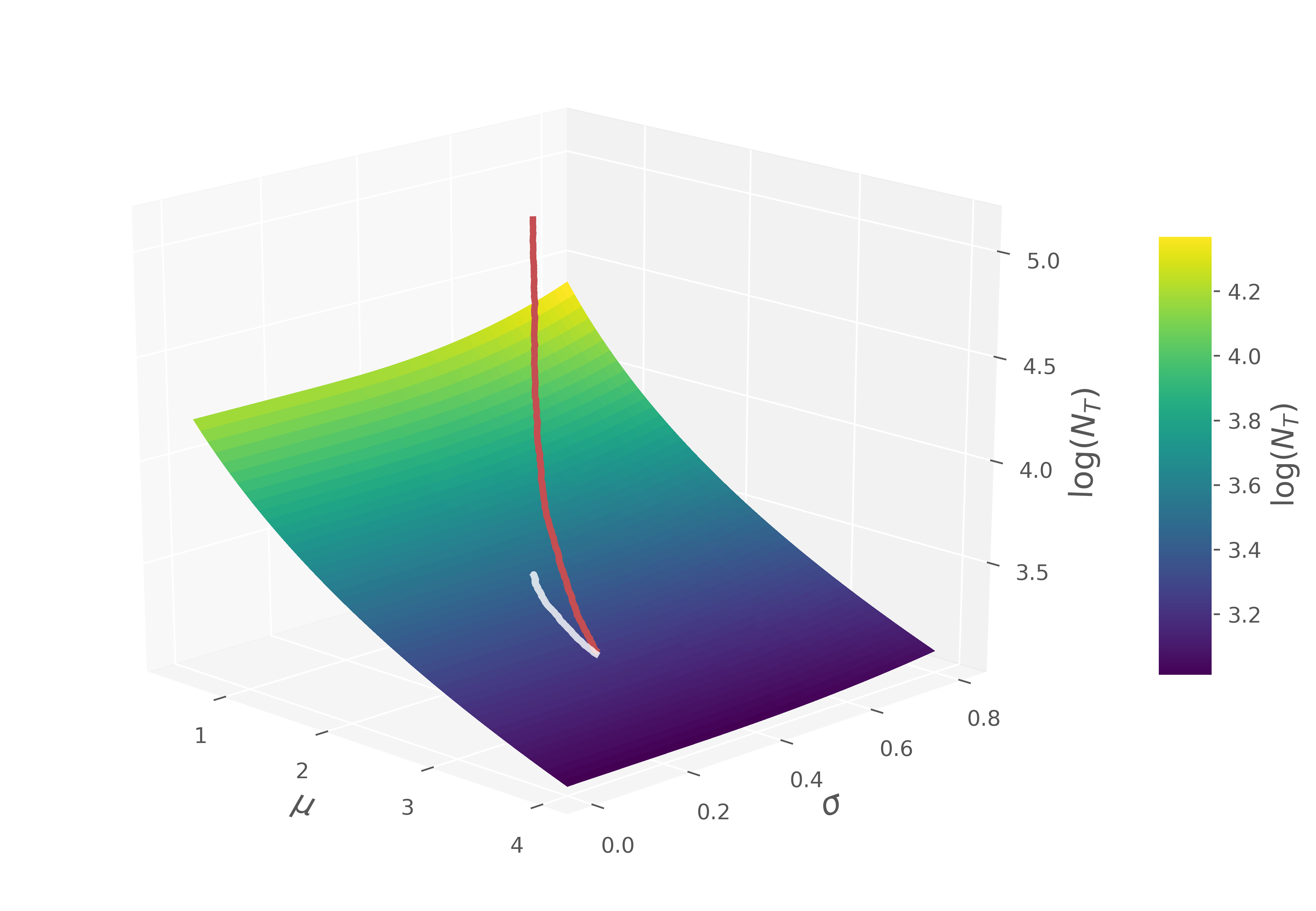}
    \caption{{\bf Purely random interactions cannot explain the evolution of total community abundance.} Variation of the interaction moments $\mu(\tau)$, $\sigma(\tau)$, and of the total abundance $\log(N_T(\tau))$ (red line) along an evolutionary trajectory. The abundance of a random interaction matrix (equation \eqref{eq:random}) with moments $\mu$, $\sigma$ (surface) is plotted for comparison.
    The white line is the predicted total abundance if the matrix of moments $\mu(\tau)$, $\sigma(\tau)$ was completely random, indicating that along the trajectory the matrix becomes progressively structured.}
    \label{fig:traj_map}
\end{figure}

Analytical results obtained for disordered communities show that for random matrices defined by equation \eqref{eq:random} the total population size $N_T$ is purely a function of $\mu$ and $\sigma$. Thus, one could envision selection as a process in which the empirical moments of $\alpha$ change across community generations but the interaction matrix remains structureless as in equation \eqref{eq:random}. The evolutionary process could then be described as climbing along the gradient of the fitness function $N_T(\mu,\sigma)$ (reproduced from \citep{buninEcologicalCommunitiesLotkaVolterra2017} in Supplementary Fig. \ref{SI:phase}). 
This, however, is not what happens: the evolutionary trajectory of the community function $N_T(\mu(\tau),\sigma(\tau))$ deviates from the gradient-climbing process predicted for a random matrix with the same moments (Fig. \ref{fig:traj_map}). Hence, the  evolutionary trajectory cannot be explained only in terms of summary statistics.  One needs to dwell on the evolution of the fine-scale properties of the interaction matrix.

The mismatch between the evolved and the corresponding random matrices in eq. \eqref{eq:random} (with the same $\mu$ and $\sigma$) can be understood by looking at the evolutionary dynamics of the  eigenvalue spectrum. 
The spectrum of the initial random interaction matrix is, in the complex plane, a circle of radius $\sigma$ centred in the origin \citep{ginibreStatisticalEnsemblesComplex1965}, plus an isolated positive eigenvalue (blue in Fig. \ref{fig:stats} B) of magnitude $\mu$. The initial effect of selection is to reduce this value. After some time, however, an isolated {\it negative} eigenvalue $\lambda$ (green in Fig. \ref{fig:stats} B and C) emerges from the circle and detaches from it linearly in time.
When this happens, the interaction matrix $\alpha$ has two components. The random component, represented by a circle of eigenvalues, changes only slightly its radius along the evolutionary trajectory. The isolated eigenvalues, on the other hand, and their associated eigenvector change on the evolutionary time scale.
At the dominant order, the {\it structure imprinted by selection} on the interaction matrix is determined by its smallest eigenvalue, that corresponds to the slowest mode of the GLV equation. Such rank-one perturbation adds to eq. \eqref{LV_eq} a global mutualistic term, which pushes towards higher abundances all species that do not go extinct. 

\begin{figure}[h]
    \centering
    \includegraphics[width=11cm]{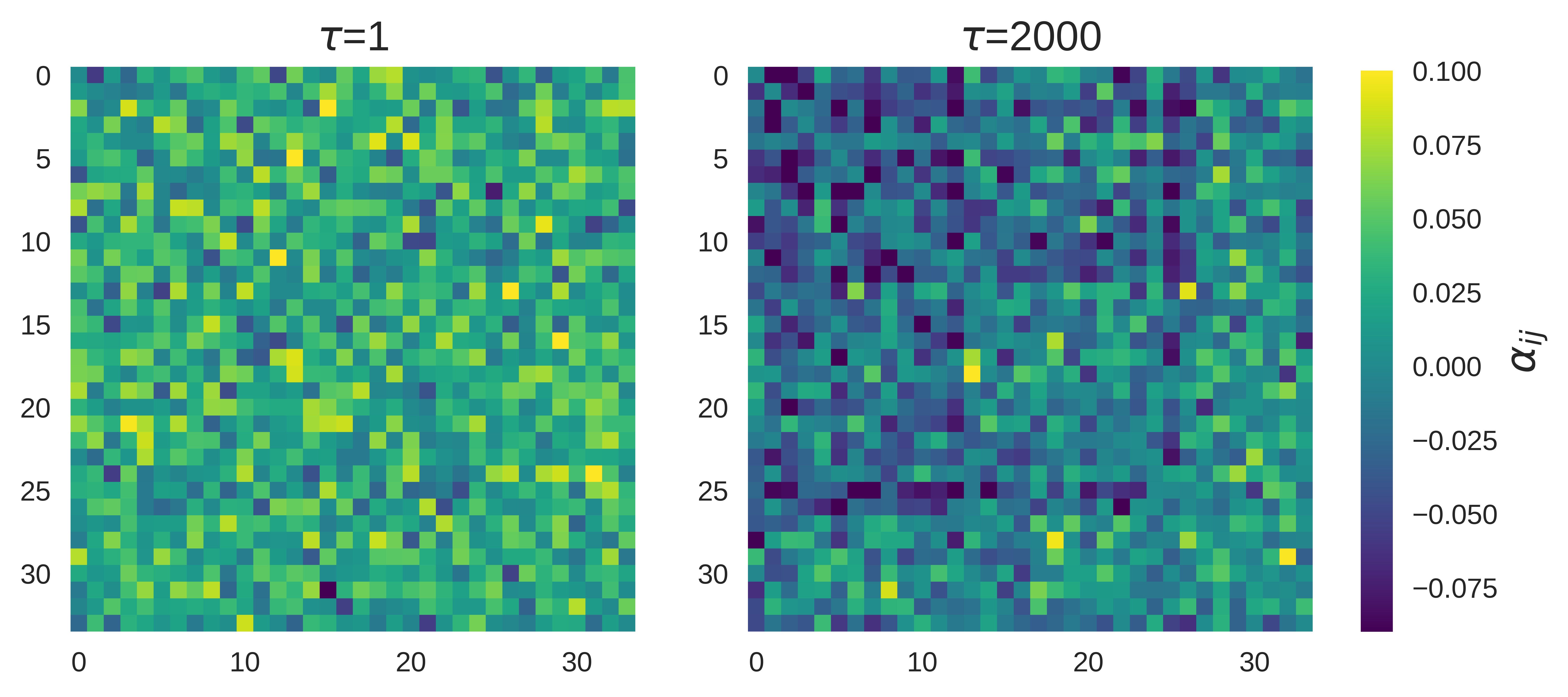}
    \caption{{\bf Evolution of the interaction matrix.} Coefficients of the interaction matrix $\alpha$ with rows and columns sorted by decreasing carrying capacities at generations 1 (left) and 2000 (right) for the same simulation as Fig. \ref{fig:abundance}. Only the species that have positive abundance at generation $2000$ are shown.}
    \label{fig:evol_alpha}
\end{figure}

The left eigenvector $\vb{q}$ associated to the outlier eigenvalue essentially retains the information on the evolved community composition, as it is strongly correlated to the equilibrium abundance vector (Supplementary Fig. \ref{SI:fig:coeffs_Nv}).
Moreover, both vectors are correlated to the vector of carrying capacities $\vb{K}$ (Supplementary Fig. \ref{SI:fig:coeffs_Nv}). As a result, species that have become more mutualistic after $2000$ generations are mostly those that initially had higher carrying capacity.
The imprinted structure that emerged along the evolutionary trajectory thus appears when the entries of $\alpha$ for early and late stages of community evolution are compared. By ordering species in terms of their carrying capacity (from larger to smaller, Fig. \ref{fig:evol_alpha}), no structure of the off-diagonal entries is visible in the ancestral matrix, while a gradient appears after selection has acted for a sufficiently long time. 

If we stop selection for increased total abundance but maintain mutations, the outlier eigenvalue goes back slowly to the circle. The evolutionary trajectory thus approaches the surface in Fig \ref{fig:traj_map}, so that the total abundance becomes purely a function of $\mu$ and $\sigma$, as it was in the ancestral community. The only trace of the elapsed community evolution remains then in the modified summary statistics of the interaction matrix, while selection is no longer directly detectable.

Simulations realised for a number of different parameter values and for other target functions (Supplementary Section \ref{SI:generalizations}) suggest that the phenomena illustrated above for asymmetric interactions ($\gamma=0$) are general.

\subsection{Analytical description of the evolutionary dynamics}

In order to understand the origin of this generality and to deduce the laws governing the evolution of the interaction matrix, we introduce a theoretical framework that links the dynamics of interactions to the parameters of the system, including those defining the experimental protocol, the target of selection and the ancestral community. \markasnew{In this part we derive equations for any community-level function $f$ and discuss how the numerical results presented above can be understood if $f$ is the total abundance.}

Given a community with interaction matrix $\alpha(\tau)$ \markasnew{(not necessarily random) and the corresponding} equilibrium abundances $\vb{N}(\tau)$ at a given generation $\tau$, we aim to characterise the interaction matrix $\alpha(\tau+1)$ of the selected offspring community -- the one that provides the largest \markasnew{community-level function $f$} at equilibrium. 

Because of mutations, the interaction matrix of each offspring communities can be written for small $\varepsilon$ as $\alpha= \alpha(\tau)+\frac{\varepsilon\sigma(\tau)}{\sqrt{S}}\eta$, with a different realisation of $\eta$ for every communities (from Eq. \eqref{mut_eq}). 
The changes of the equilibrium abundances induced by a modification of the interaction matrix are mathematically equivalent to those obtained after small random perturbations of the carrying capacities $ \delta \vb{K} = -\frac{\varepsilon\sigma(\tau)}{\sqrt{S}}\;\eta\; \vb{N}(\tau)$. 
Linear response theory provides the corresponding change induced on the equilibrium abundances: 
\begin{equation}\label{eq:perturb}
\delta \vb{N} =\chi(\tau)\delta\,\vb{K}=-\frac{\varepsilon\,\sigma(\tau)}{\sqrt{S}} \chi\,\eta \vb{N}(\tau)
\end{equation}
with $\chi_{ij}=\frac{\partial N_i}{\partial K_j}$ the stability matrix. This matrix measures the effect of a small change in the carrying capacities on the abundances at equilibrium and depends non-linearly on the interaction matrix \markasnew{(see Supplementary section {\ref{SI:first_order}})}.

If $f(\vb{N})$ is the community-level function that one wants to maximise, its value at equilibrium for each community is thus equal to the value at generation $\tau$, plus a small random change that derives from the variation in the abundances. Using eq. \eqref{eq:perturb}, this random contribution can be written as:
\begin{equation}
    \delta f = \grad f(\vb{N})\cdot\delta\vb{N}=-\frac{\varepsilon\,\sigma(\tau)}{\sqrt{S}}\,\grad f(\vb{N})\cdot\chi(\tau)\,\eta \,\vb{N}.
\end{equation}
The largest improvement of the function that will be realised in the pool of mutated communities can therefore be identified as the largest among the $n$ independent random variables \markasnew{$\delta f$} \citep{gumbelStatisticsExtremes2004}. 
\markasnew{In the Supplementary section {\ref{SI:max}}, we derive the statistics of such largest contribution $\delta f_\star$, which is related to a random variable $M_n$ following the law of the maximum value of $n$ independent Gaussian variables (see the distribution of $M_n$ in Supplementary Fig.{\ref{SI:distrib})}. The community-level function of the selected community is then simply the sum of its value at the previous generation and of this largest contribution: $f(\tau+1)=f(\tau)+\delta f_\star$. As we explain in detail in the Supplementary section {\ref{SI:section:mut_sel}}, this gives the expression:}
\begin{equation}\label{eq:N_rec}
    f(\tau+1)=f(\tau)+M_n(\tau)\,\frac{\varepsilon\sigma(\tau)}{\sqrt{S}}\,\| \vb{v}(\tau)\|\,\|\vb{N}(\tau)\|,
\end{equation}
where: 
\begin{equation}
\vb{v}(\tau)=\chi^{\top}(\tau)\,\grad f(\vb{N}(\tau))
\end{equation}
is \markasnew{a vector representing how the function $f$ measured at equilibrium changes when we vary the carrying capacities: $v_i = \frac{\partial f(\vb{N})}{\partial K_i}$ \footnote{The dependency on the carrying capacities is mediated by the equilibrium abundances.}. This "sensitivity vector" depends both on the interaction matrix and on the function $f$.} 

In the specific case of $f(\vb{N})$ being the total abundance $N_T=\vb{N}\cdot \vb{1}$, we have $\vb{v}(\tau)=\chi^{\top}(\tau)\,\vb{1}$. 

Equation \eqref{eq:N_rec} implies that the community function increases on average along an evolutionary trajectory, as the product of the norms is always positive and $M_n$ has a positive expected value for $n>1$.  When the number of communities is too small, it can also transiently decrease, thus breaking the alignment between selection and community response.
For large $n$ the distribution of $M_n$ concentrates around its mean for $n\gg 1$, thus making this event very unlikely. 

Changes in \markasnew{the community-level function} are ultimately based on the evolution of the interaction matrix. 
As detailed in the SI, its change across one collective generation can be decomposed in a directional term -- contributing to the evolution of \markasnew{$f$} -- and its complement $B_{ij}$, that acts as a random fluctuation\footnote{Formally, this amounts to decomposing $\eta_{ij}$ in two parts: one along the direction (in the space of the $S^2$ indices) $\frac{v_i(\tau)}{\| \vb{v}(\tau)\|} \frac{N_j(\tau)}{\|\vb{N}(\tau)\|}$, and a remainder $B_{ij}=\eta_{ij}-\frac{v_i(\tau)}{\| \vb{v}(\tau)\|}\frac{N_j(\tau)}{\|\vb{N}(\tau)\|}\sum_{k,l}\frac{v_k(\tau)}{\| \vb{v}(\tau)\|} \eta_{kl} \frac{N_l(\tau)}{\|\vb{N}(\tau)\|}$}.
The interaction between any two species $i$ and $j$ thus evolves according to:
\begin{equation}\label{eq:alpha_rec}
    \alpha_{ij}(\tau +1) = \alpha_{ij}(\tau)-\frac{\varepsilon \sigma(\tau)}{\sqrt{S}}\left(\,M_n(\tau)\frac{v_i(\tau)}{\| \vb{v}(\tau)\|} \, \frac{N_j(\tau)}{\|\vb{N}(\tau)\|}+B_{ij}\right),
\end{equation}
which encodes the evolutionary dynamics of the interaction matrix. 
This expression (generalised for $\gamma \neq 0$ in the Supplementary equation \eqref{SI:eq:alpha_rec}) has a simple interpretation: among the random mutations of the interaction matrix, only matter those in the special "direction" \markasnew{$v_iN_j$, that combines the sensitivity of the community function $f$ and the equilibrium abundances.} The selected community is the one having the largest random Gaussian contribution associated to such direction. 

\markasnew{A direct consequence of equation {\eqref{eq:alpha_rec}} is that the most abundant species will experience greater changes in the impact they have on other species ($\alpha_{ij}$), but these changes can be of any sign, depending on the sensitivity with respect to the other species involved ($v_i$). Conversely, species with the most positive impact on the function (those with larger $v_i$) will face a greater reduction in the effects that any other species has on them.}

\markasnew{It is interesting to note that in equations {\eqref{eq:N_rec}} and {\eqref{eq:alpha_rec}}, the community function only appears through the vector $\vb{v}$.}
\markasnew{Because this vector depends both on the interaction matrix of the community and on the function $f$,} different communities will have different responses to the same target functions, and, vice versa, the same community may react differently to selection depending on the function it is selected for. 
Matching selection target and community structure is therefore determinant for speeding up evolution, and could be improved by preliminary tests of the community response to perturbations.
In the special case when $f(\vb{N})$ does not depend on $\vb{N}$, instead, selection at the community level is neutral (any community composition is equivalent), and interactions evolve by drift, driven by the random term $B_{ij}$.

Equations \eqref{eq:N_rec} and \eqref{eq:alpha_rec} apply to any initial interaction matrix (not only random ones) and allow us to draw general conclusions, which we spell out below, on how speed and direction of evolutionary change depend on the numerous parameters of the system. 

As could be intuited, evolution is faster when selection screens a larger number of communities, since the expected value $\overline{M_n}$ is an increasing function of $n$. 
When only one community is considered, on the other hand, the total abundance and the interaction matrix undergo unbiased stochastic changes (see Supplementary section \ref{SI:random_walk}), as $M_1$ is Gaussian with zero mean. Under these conditions, collective functions cannot be selected and evolve by community-level drift. 
However, increasing the number of communities may not always be the key to success. The growth of $\overline{M_n}$ with $n$, indeed, scales as $\sqrt{\log(n)}$, which increases slowly for large $n$, so that transition to high community throughput may be of little avail to speed up evolution. 

Other parameters can be changed so as to improve the efficacy of community selection. The variation of the interaction matrix, thus of the selected function, across one community generation occurs on a time scale $dt=\varepsilon/\sqrt{S}$. Thus, in our model, evolution has faster pace in communities with a smaller number of species and for larger mutational steps. This is however linked to the choices we made for the initial interactions and for the mutations, and we expect that different assumptions may lead to other scaling laws.

The non-linear recursive matrix \markasnew{equation} (\ref{eq:alpha_rec}) cannot be solved in general, so as to exactly predict how the ancestral community changes along an evolutionary trajectory. Whatever the exact change, however, it shows that selection will -- sooner or later -- result in a perturbation of rank one of the interaction matrix, which translates into the addition of a \markasnew{rank one} term in the original GLV equation \eqref{LV_eq}. This term, as well as the whole dynamics, can be explicitly computed \markasnew{when the selected function is the total abundance and} in the limiting case of small variability of interactions $\sigma \ll 1$ (see Supplementary section \ref{SI:section:small_sig}). The effect of selection is here simply to \markasnew{decrease all the interactions by the same amount $\varepsilon M_n\sigma \, \sqrt{1+\gamma}  / \sqrt{S}$ at each generation} (Supplementary equation \eqref{SI:eq:dmudt}), so that they progressively become more mutualistic, up to the point where the selected function diverges. 

The exactly solvable, approximate solution moreover highlights the role of the symmetry of interactions in the efficiency of artificial selection. 
Indeed, evolution is fastest when $\gamma=1$, \textit{i.e.} for competitive or mutualistic interactions. On the contrary, when interactions are antagonistic, such as predator-prey or host-parasite (in the extreme cases, $\gamma=-1$), very little improvement has to be expected when applying selection for increased abundance. Intuitively, this is because variations in the abundance of the two interacting partners are negatively correlated, so that their global effects cancel out.

A remarkable feature in the evolution of interactions is, as illustrated earlier by numerical simulations (Fig. \ref{fig:stats} C), the existence of a finite time where the matrix transitions from random to acquiring new structure in the spectrum. 
The emergence of an isolated eigenvalue from the random circle when a strong enough rank one term is added to a random matrix is known in statistical physics and signal processing as BBP phase transition \citep{baikPhaseTransitionLargest2005a}. The transition we find has exactly the same properties of a BBP one. However, unlike this transition, in our case the moment when this will happen and the associated eigenvalues cannot be predicted because of the random Gaussian contributions $B_{ij}$. They do not modify the initial random structure, but they change in time whenever interaction diversity is not vanishing\footnote{In mathematical terms, the difference with the standard BBP transition is that instead of adding always the same rank-one contribution, in eq. \eqref{eq:alpha_rec} the added rank one contribution changes with time making the analysis challenging.}. As a consequence, the composition of the evolved community at the moment of the transition is not unique despite the clear statistical signature.

How can one then distinguish communities that have undergone the transition, whose composition is constrained by the alignment to the eigenvector associated to the isolated eigenvalue? If all the pairwise interaction coefficients were know, it would be straightforward to compute the spectrum of the interaction matrix. But in large communities the estimation of such coefficients is very time-consuming even for one point in time, and complete time series may be hardly accessible. We thus explored the possibility of inferring the transition time from observables that are more readily accessible.
By leveraging the maximum entropy method developed in \citep{barbierFingerprintsHighDimensionalCoexistence2021} we can characterise the class of the most likely interaction matrices, knowing the mean interaction $\mu$, the equilibrium abundances $\vb{N}$ and the carrying capacity vector $\vb{K}$. For large $S$, these matrices can be written (see Supplementary section \ref{SI:section:max_ent}):
\begin{equation}\label{eq:random-me}
    \tilde \alpha_{ij}(\tau)=\frac{l_i(\tau) N_j(\tau)}{\sum_k N_k^2(\tau)}+\sigma(\tau) z_{ij},
\end{equation}
where $l_i(\tau)=K_i-N_i(\tau)-\mu(\tau)N_T(\tau)/S$ and $z_{ij}$ is a Gaussian random matrix of zero mean and unit variance. 
\footnote{Here we are only interested in the spectrum and the isolated eigenvalue.}
When $\|\vb{l}(\tau)\|>\sigma(\tau)\| \vb{N}(\tau)\|$, this matrix undergoes a BBP transition, and an isolated eigenvalue emerges from a circle of radius $\sigma(\tau)$. 
We find that this maximum entropy method provides a very accurate description of the evolutionary dynamics of the interaction matrix. In fact,  
both the isolated eigenvalue and the associated eigenvectors of the maximum entropy inferred matrix agree remarkably well with the actual ones, all along the evolutionary trajectory (Supplementary Fig. \ref{SI:fig:max_ent}). These results suggest that it may be possible to detect the low-dimensional structure imprinted by artificial selection of species-rich communities even in experimental settings. 

\section{Discussion}
This study is devoted to identifying key and general features of the evolutionary dynamics in species-rich communities under a scheme that is commonly used for artificial selection of collective functions.
We showed that the interaction matrix evolves in response to selection for total abundance, and that it results generically in interspecific interactions becoming progressively less competitive. We interpret this as the evolution of facilitation, similar to what was observed in a two-species model \citep{doulcierEcoevolutionaryDynamicsNested2020}.
At the same time as the average strength of interspecific interactions decreases, they become more variable.
Notably, the evolutionary process imprints a structure on the interaction matrix. The key to this structure is an isolated eigenvalue, which emerges as a 'collective mode' that positively impacts the abundances of all species. In the analytical description, this corresponds to a \markasnew{rank-one} perturbation of the interaction matrix, that otherwise retains its original, disordered nature.

The emergence of structure in the form of a low-rank perturbation is not specific to selection acting on total abundance, but is predicted to hold for any function of the abundances at equilibrium.
This seems indeed to be a general feature of systems with a high number of degrees of freedom whose interactions are dynamically adjusted for achieving a specific collective goal, such as a lower ground state energy in spin-glasses and learning in neural networks \citep{penney1993coupled,saxe2019mathematical, schuesslerInterplayRandomnessStructure2021}. 
The ubiquity of this phenomenon raises the question if and when selection can produce the emergence of more complex structures, for instance the emergence of several distinct dominant eigenvalues. 

\markasnew{These findings have valuable implications for the formulation of models that incorporate further biological realism. In particular, they suggest that, from the point of view of community-level selection, relevant modifications of basic disordered models are those that produce low-rank terms in the interaction matrix. These terms may compete with selection acting at the collective level, and stir the evolutionary path of the community.}

We chose to analyse an idealised model in order to achieve analytical tractability.
If disordered models are certainly an oversimplification of real communities, they have the double advantage of not relying on detailed descriptions of the community, and of providing null expectations for how collective properties would evolve in the absence of species-level constrains. In fact, the actual strength of ecological interactions is unknown in most microbial communities. Statistical approaches, that represent interactions in terms of a few key parameters, can then be a valuable method for identifying general prescriptions relevant in experiments \citep{barbierGenericAssemblyPatterns2018, hu_emergent_2022}. 

The model we have studied may be extended in several meaningful ways.
Instead of modelling species interactions through direct effects, one could include explicitly the resources that are consumed or exchanged \citep{williamsArtificialSelectionSimulated2007,cuiDiverseCommunitiesBehave2021,changEngineeringComplexCommunities2021a}. Given the equivalence of the Lotka-Volterra and MacArthur models when resource dynamics is much faster than the ecological one, we expect that our main results qualitatively hold in this case too. However, a formulation in terms of resource consumption would connect theoretical results to experiments exploring the metabolic foundations of ecological interactions in microbial communities \citep{faustMicrobialInteractionsNetworks2012a,estrelaMultiReplicatedEnrichmentCommunities2021}. Especially, this may guide the choice of more realistic interaction matrices, such as sparse networks \citep{marcus_local_2022, buninDirectionalityCommunitylevelSelection2021}, or networks with empirical biases \citep{machadoPolarizationMicrobialCommunities2021}.

Even maintaining random direct interactions, the model we considered could be explored in regimes where the perturbative approach is expected to break down. This would occur for instance when the ecological dynamics of the community does not attain an equilibrium because of transients \citep{changEngineeringComplexCommunities2021a}, stochastic demographic fluctuations \citep{altieri2021properties} or chaotic population dynamics \citep{rogers_chaos_2022, buninEcologicalCommunitiesLotkaVolterra2017, biroliMarginallyStableEquilibria2018}. All these processes may reduce the heritability of the community function and thus alter the evolutionary trajectory.

Finally, consistent with the idea that communities are Darwinian individuals \citep{godfrey2009darwinian}, we chose mutations that would provide unbiased community-level variation in the target function. Such assumption allowed us to develop a null model that is independent of the details of the underlying community interactions. Collective-level mutations can be thought of as the result of multiple changes in species interactions that occurred during the lifetime of a community. More detailed descriptions of how sequential species-level mutations give rise to variation of the interaction matrix at the time of community reproduction -- that is when the function is evaluated -- are worth studying and may prove necessary for specific applications (they could provide additional constraints, as observed for simpler models \citep{doulcierEcoevolutionaryDynamicsNested2020}).
Furthermore, the model could be extended so as to include mutations of intra-species interactions via changes of the carrying capacities or speciation events that would increase diversity.

Communities are increasingly conceived as coherent units that provide collective-level functions, to the point to be attributed the status of 'organisms' \citep{wilsonRevivingSuperorganism1989,loreauEcosystemSuperorganismCollection2020}. If this view can reflect the way ecological interactions produce a given population structure \citep{liautaudSuperorganismsLooseCollections2019}, it can go as far as identifying communities as full-fledged evolutionary units. In the latter case, how they are 'scaffolded' by physical compartmentalisation and the establishment of community-level lineages, is all-important in determining the action of natural selection at the level of communities \citep{demonteNascentMulticellularLife2014,blackEcologicalScaffoldingEvolution2020a}. We have modelled here the protocol commonly used in experiments of artificial selection \citep{arias-sanchezArtificiallySelectingMicrobial2019,sanchezDirectedEvolutionMicrobial2021}. Considering that the collective level is the true center of interest for this process, moreover, we described mutations only for their effect on the community-level function under selection. Nested levels of reproducing units are widespread in the hierarchy of living beings. Our results might thus be relevant whenever selection on high-level functions bestows a structure on the interaction among heterogeneous constituent units, and contribute to understanding how integration across levels of organisation is achieved.

\section*{Acknowledgments}
We thank Matthieu Barbier for insightful discussions.
This research was partially supported by a grant from the Simons Foundation (N. 454935 Giulio Biroli). SDM was supported by the French Government under the program Investissements d’Avenir (ANR-10-LABX-54 MEMOLIFE and ANR-11-IDEX- 0001-02PSL).

\bibliographystyle{elsarticle-harv}
\bibliography{biblio}

\begin{thebibliography}{49}
\expandafter\ifx\csname natexlab\endcsname\relax\def\natexlab#1{#1}\fi
\providecommand{\url}[1]{\texttt{#1}}
\providecommand{\href}[2]{#2}
\providecommand{\path}[1]{#1}
\providecommand{\DOIprefix}{doi:}
\providecommand{\ArXivprefix}{arXiv:}
\providecommand{\URLprefix}{URL: }
\providecommand{\Pubmedprefix}{pmid:}
\providecommand{\doi}[1]{\href{http://dx.doi.org/#1}{\path{#1}}}
\providecommand{\Pubmed}[1]{\href{pmid:#1}{\path{#1}}}
\providecommand{\bibinfo}[2]{#2}
\ifx\xfnm\relax \def\xfnm[#1]{\unskip,\space#1}\fi
%Type = Incollection
\bibitem[{Alessina(2020)}]{dobson_going_2020}
\bibinfo{author}{Alessina, S.}, \bibinfo{year}{2020}.
\newblock \bibinfo{title}{Going {Big}}, in: \bibinfo{editor}{Dobson, A.},
  \bibinfo{editor}{Tilman, D.}, \bibinfo{editor}{Holt, R.D.} (Eds.),
  \bibinfo{booktitle}{Unsolved {Problems} in {Ecology}}.
  \bibinfo{publisher}{Princeton University Press}.
\newblock \bibinfo{note}{Google-Books-ID: 4Q3MDwAAQBAJ}.
%Type = Article
\bibitem[{Altieri et~al.(2021)Altieri, Roy, Cammarota and
  Biroli}]{altieri2021properties}
\bibinfo{author}{Altieri, A.}, \bibinfo{author}{Roy, F.},
  \bibinfo{author}{Cammarota, C.}, \bibinfo{author}{Biroli, G.},
  \bibinfo{year}{2021}.
\newblock \bibinfo{title}{Properties of equilibria and glassy phases of the
  random lotka-volterra model with demographic noise}.
\newblock \bibinfo{journal}{Physical Review Letters} \bibinfo{volume}{126},
  \bibinfo{pages}{258301}.
%Type = Article
\bibitem[{{Arias-S{\'a}nchez} et~al.(2019){Arias-S{\'a}nchez}, Vessman and
  Mitri}]{arias-sanchezArtificiallySelectingMicrobial2019}
\bibinfo{author}{{Arias-S{\'a}nchez}, F.I.}, \bibinfo{author}{Vessman, B.},
  \bibinfo{author}{Mitri, S.}, \bibinfo{year}{2019}.
\newblock \bibinfo{title}{Artificially selecting microbial communities: {{If}}
  we can breed dogs, why not microbiomes?}
\newblock \bibinfo{journal}{PLOS Biology} \bibinfo{volume}{17},
  \bibinfo{pages}{e3000356}.
\newblock \DOIprefix\doi{10.1371/journal.pbio.3000356}.
%Type = Article
\bibitem[{Baik et~al.(2005)Baik, Arous and
  P{\'e}ch{\'e}}]{baikPhaseTransitionLargest2005a}
\bibinfo{author}{Baik, J.}, \bibinfo{author}{Arous, G.B.},
  \bibinfo{author}{P{\'e}ch{\'e}, S.}, \bibinfo{year}{2005}.
\newblock \bibinfo{title}{Phase transition of the largest eigenvalue for
  nonnull complex sample covariance matrices}.
\newblock \bibinfo{journal}{Annals of Probability} \bibinfo{volume}{33},
  \bibinfo{pages}{1643--1697}.
\newblock \DOIprefix\doi{10.1214/009117905000000233}.
%Type = Article
\bibitem[{Barbier et~al.(2018)Barbier, Arnoldi, Bunin and
  Loreau}]{barbierGenericAssemblyPatterns2018}
\bibinfo{author}{Barbier, M.}, \bibinfo{author}{Arnoldi, J.F.},
  \bibinfo{author}{Bunin, G.}, \bibinfo{author}{Loreau, M.},
  \bibinfo{year}{2018}.
\newblock \bibinfo{title}{Generic assembly patterns in complex ecological
  communities}.
\newblock \bibinfo{journal}{Proceedings of the National Academy of Sciences}
  \bibinfo{volume}{115}, \bibinfo{pages}{2156--2161}.
\newblock \DOIprefix\doi{10.1073/pnas.1710352115}.
%Type = Article
\bibitem[{Barbier et~al.(2021)Barbier, {de Mazancourt}, Loreau and
  Bunin}]{barbierFingerprintsHighDimensionalCoexistence2021}
\bibinfo{author}{Barbier, M.}, \bibinfo{author}{{de Mazancourt}, C.},
  \bibinfo{author}{Loreau, M.}, \bibinfo{author}{Bunin, G.},
  \bibinfo{year}{2021}.
\newblock \bibinfo{title}{Fingerprints of {{High-Dimensional Coexistence}} in
  {{Complex Ecosystems}}}.
\newblock \bibinfo{journal}{Physical Review X} \bibinfo{volume}{11},
  \bibinfo{pages}{011009}.
\newblock \DOIprefix\doi{10.1103/PhysRevX.11.011009}.
%Type = Article
\bibitem[{van~den Berg et~al.(2022)van~den Berg, Machado, Santos, Rocha,
  Chacón, Harcombe, Mitri and Patil}]{van_den_berg_ecological_2022}
\bibinfo{author}{van~den Berg, N.I.}, \bibinfo{author}{Machado, D.},
  \bibinfo{author}{Santos, S.}, \bibinfo{author}{Rocha, I.},
  \bibinfo{author}{Chacón, J.}, \bibinfo{author}{Harcombe, W.},
  \bibinfo{author}{Mitri, S.}, \bibinfo{author}{Patil, K.R.},
  \bibinfo{year}{2022}.
\newblock \bibinfo{title}{Ecological modelling approaches for predicting
  emergent properties in microbial communities}.
\newblock \bibinfo{journal}{Nature Ecology \& Evolution} \bibinfo{volume}{6},
  \bibinfo{pages}{855--865}.
\newblock \URLprefix \url{https://www.nature.com/articles/s41559-022-01746-7},
  \DOIprefix\doi{10.1038/s41559-022-01746-7}.
%Type = Article
\bibitem[{Biroli et~al.(2018)Biroli, Bunin and
  Cammarota}]{biroliMarginallyStableEquilibria2018}
\bibinfo{author}{Biroli, G.}, \bibinfo{author}{Bunin, G.},
  \bibinfo{author}{Cammarota, C.}, \bibinfo{year}{2018}.
\newblock \bibinfo{title}{Marginally stable equilibria in critical ecosystems}.
\newblock \bibinfo{journal}{New Journal of Physics} \bibinfo{volume}{20},
  \bibinfo{pages}{083051}.
\newblock \DOIprefix\doi{10.1088/1367-2630/aada58}.
%Type = Article
\bibitem[{Black et~al.(2020)Black, Bourrat and
  Rainey}]{blackEcologicalScaffoldingEvolution2020a}
\bibinfo{author}{Black, A.J.}, \bibinfo{author}{Bourrat, P.},
  \bibinfo{author}{Rainey, P.B.}, \bibinfo{year}{2020}.
\newblock \bibinfo{title}{Ecological scaffolding and the evolution of
  individuality}.
\newblock \bibinfo{journal}{Nature Ecology \& Evolution} \bibinfo{volume}{4},
  \bibinfo{pages}{426--436}.
\newblock \DOIprefix\doi{10.1038/s41559-019-1086-9}.
%Type = Article
\bibitem[{Bunin(2017)}]{buninEcologicalCommunitiesLotkaVolterra2017}
\bibinfo{author}{Bunin, G.}, \bibinfo{year}{2017}.
\newblock \bibinfo{title}{Ecological communities with {{Lotka}}-{{Volterra}}
  dynamics}.
\newblock \bibinfo{journal}{Physical Review E} \bibinfo{volume}{95},
  \bibinfo{pages}{042414}.
\newblock \DOIprefix\doi{10.1103/PhysRevE.95.042414}.
%Type = Article
\bibitem[{Bunin(2021)}]{buninDirectionalityCommunitylevelSelection2021}
\bibinfo{author}{Bunin, G.}, \bibinfo{year}{2021}.
\newblock \bibinfo{title}{Directionality and community-level selection}.
\newblock \bibinfo{journal}{Oikos} \bibinfo{volume}{130},
  \bibinfo{pages}{489--500}.
\newblock \DOIprefix\doi{10.1111/oik.07214}.
%Type = Article
\bibitem[{Chang et~al.(2021)Chang, Vila, Bender, Li, Mankowski, Bassette,
  Borden, Golfier, Sanchez, Waymack, Zhu, {Diaz-Colunga}, Estrela,
  {Rebolleda-Gomez} and Sanchez}]{changEngineeringComplexCommunities2021a}
\bibinfo{author}{Chang, C.Y.}, \bibinfo{author}{Vila, J.C.C.},
  \bibinfo{author}{Bender, M.}, \bibinfo{author}{Li, R.},
  \bibinfo{author}{Mankowski, M.C.}, \bibinfo{author}{Bassette, M.},
  \bibinfo{author}{Borden, J.}, \bibinfo{author}{Golfier, S.},
  \bibinfo{author}{Sanchez, P.G.L.}, \bibinfo{author}{Waymack, R.},
  \bibinfo{author}{Zhu, X.}, \bibinfo{author}{{Diaz-Colunga}, J.},
  \bibinfo{author}{Estrela, S.}, \bibinfo{author}{{Rebolleda-Gomez}, M.},
  \bibinfo{author}{Sanchez, A.}, \bibinfo{year}{2021}.
\newblock \bibinfo{title}{Engineering complex communities by directed
  evolution}.
\newblock \bibinfo{journal}{Nature Ecology \& Evolution} \bibinfo{volume}{5},
  \bibinfo{pages}{1011--1023}.
\newblock \DOIprefix\doi{10.1038/s41559-021-01457-5}.
%Type = Article
\bibitem[{Cui et~al.(2021)Cui, Marsland and
  Mehta}]{cuiDiverseCommunitiesBehave2021}
\bibinfo{author}{Cui, W.}, \bibinfo{author}{Marsland, R.},
  \bibinfo{author}{Mehta, P.}, \bibinfo{year}{2021}.
\newblock \bibinfo{title}{Diverse communities behave like typical random
  ecosystems}.
\newblock \bibinfo{journal}{Physical Review E} \bibinfo{volume}{104},
  \bibinfo{pages}{034416}.
\newblock \DOIprefix\doi{10.1103/PhysRevE.104.034416}.
%Type = Article
\bibitem[{De~Monte(2021)}]{demonteEcologicalRecipesSelecting2021}
\bibinfo{author}{De~Monte, S.}, \bibinfo{year}{2021}.
\newblock \bibinfo{title}{Ecological recipes for selecting community function}.
\newblock \bibinfo{journal}{Nature Ecology \& Evolution} \bibinfo{volume}{5},
  \bibinfo{pages}{894--895}.
\newblock \DOIprefix\doi{10.1038/s41559-021-01467-3}.
%Type = Article
\bibitem[{De~Monte and Rainey(2014)}]{demonteNascentMulticellularLife2014}
\bibinfo{author}{De~Monte, S.}, \bibinfo{author}{Rainey, P.B.},
  \bibinfo{year}{2014}.
\newblock \bibinfo{title}{Nascent multicellular life and the emergence of
  individuality}.
\newblock \bibinfo{journal}{Journal of Biosciences} \bibinfo{volume}{39},
  \bibinfo{pages}{237--248}.
\newblock \DOIprefix\doi{10.1007/s12038-014-9420-5}.
%Type = Article
\bibitem[{Doulcier et~al.(2020)Doulcier, Lambert, De~Monte and
  Rainey}]{doulcierEcoevolutionaryDynamicsNested2020}
\bibinfo{author}{Doulcier, G.}, \bibinfo{author}{Lambert, A.},
  \bibinfo{author}{De~Monte, S.}, \bibinfo{author}{Rainey, P.B.},
  \bibinfo{year}{2020}.
\newblock \bibinfo{title}{Eco-evolutionary dynamics of nested {{Darwinian}}
  populations and the emergence of community-level heredity}.
\newblock \bibinfo{journal}{eLife} \bibinfo{volume}{9},
  \bibinfo{pages}{e53433}.
\newblock \DOIprefix\doi{10.7554/eLife.53433}.
%Type = Article
\bibitem[{Estrela et~al.(2021)Estrela, S{\'a}nchez and
  {Rebolleda-G{\'o}mez}}]{estrelaMultiReplicatedEnrichmentCommunities2021}
\bibinfo{author}{Estrela, S.}, \bibinfo{author}{S{\'a}nchez, {\'A}.},
  \bibinfo{author}{{Rebolleda-G{\'o}mez}, M.}, \bibinfo{year}{2021}.
\newblock \bibinfo{title}{Multi-{{Replicated Enrichment Communities}} as a
  {{Model System}} in {{Microbial Ecology}}}.
\newblock \bibinfo{journal}{Frontiers in Microbiology} \bibinfo{volume}{12},
  \bibinfo{pages}{760}.
\newblock \DOIprefix\doi{10.3389/fmicb.2021.657467}.
%Type = Article
\bibitem[{Faust and Raes(2012)}]{faustMicrobialInteractionsNetworks2012a}
\bibinfo{author}{Faust, K.}, \bibinfo{author}{Raes, J.}, \bibinfo{year}{2012}.
\newblock \bibinfo{title}{Microbial interactions: From networks to models}.
\newblock \bibinfo{journal}{Nature Reviews Microbiology} \bibinfo{volume}{10},
  \bibinfo{pages}{538--550}.
\newblock \DOIprefix\doi{10.1038/nrmicro2832}.
%Type = Article
\bibitem[{Fujimura et~al.(2010)Fujimura, Slusher, Cabana and
  Lynch}]{fujimuraRoleGutMicrobiota2010}
\bibinfo{author}{Fujimura, K.E.}, \bibinfo{author}{Slusher, N.A.},
  \bibinfo{author}{Cabana, M.D.}, \bibinfo{author}{Lynch, S.V.},
  \bibinfo{year}{2010}.
\newblock \bibinfo{title}{Role of the gut microbiota in defining human health}.
\newblock \bibinfo{journal}{Expert Review of Anti-infective Therapy}
  \bibinfo{volume}{8}, \bibinfo{pages}{435--454}.
\newblock \DOIprefix\doi{10.1586/eri.10.14}.
%Type = Article
\bibitem[{Garcia~Lorenzana and Altieri(2022)}]{garcia2021well}
\bibinfo{author}{Garcia~Lorenzana, G.}, \bibinfo{author}{Altieri, A.},
  \bibinfo{year}{2022}.
\newblock \bibinfo{title}{Well-mixed lotka-volterra model with random strongly
  competitive interactions}.
\newblock \bibinfo{journal}{Physical Review E} \bibinfo{volume}{105},
  \bibinfo{pages}{024307}.
\newblock \URLprefix
  \url{https://link.aps.org/doi/10.1103/PhysRevE.105.024307},
  \DOIprefix\doi{10.1103/PhysRevE.105.024307}.
%Type = Article
\bibitem[{Ginibre(1965)}]{ginibreStatisticalEnsemblesComplex1965}
\bibinfo{author}{Ginibre, J.}, \bibinfo{year}{1965}.
\newblock \bibinfo{title}{Statistical {{Ensembles}} of {{Complex}},
  {{Quaternion}}, and {{Real Matrices}}}.
\newblock \bibinfo{journal}{Journal of Mathematical Physics}
  \bibinfo{volume}{6}, \bibinfo{pages}{440--449}.
\newblock \DOIprefix\doi{10.1063/1.1704292}.
%Type = Book
\bibitem[{Godfrey-Smith(2009)}]{godfrey2009darwinian}
\bibinfo{author}{Godfrey-Smith, P.}, \bibinfo{year}{2009}.
\newblock \bibinfo{title}{Darwinian populations and natural selection}.
\newblock \bibinfo{publisher}{Oxford University Press}.
%Type = Book
\bibitem[{Gumbel(2004)}]{gumbelStatisticsExtremes2004}
\bibinfo{author}{Gumbel, E.J.}, \bibinfo{year}{2004}.
\newblock \bibinfo{title}{Statistics of {{Extremes}}}.
\newblock \bibinfo{publisher}{{Courier Corporation}}.
%Type = Article
\bibitem[{Hansen et~al.(2007)Hansen, Rainey, Haagensen and
  Molin}]{hansen_evolution_2007}
\bibinfo{author}{Hansen, S.K.}, \bibinfo{author}{Rainey, P.B.},
  \bibinfo{author}{Haagensen, J.A.J.}, \bibinfo{author}{Molin, S.},
  \bibinfo{year}{2007}.
\newblock \bibinfo{title}{Evolution of species interactions in a biofilm
  community}.
\newblock \bibinfo{journal}{Nature} \bibinfo{volume}{445},
  \bibinfo{pages}{533--536}.
\newblock \URLprefix \url{https://www.nature.com/articles/nature05514},
  \DOIprefix\doi{10.1038/nature05514}.
%Type = Article
\bibitem[{Hooper et~al.(2005)Hooper, Chapin, Ewel, Hector, Inchausti, Lavorel,
  Lawton, Lodge, Loreau, Naeem, Schmid, Set{\"a}l{\"a}, Symstad, Vandermeer and
  Wardle}]{hooperEffectsBiodiversityEcosystem2005}
\bibinfo{author}{Hooper, D.U.}, \bibinfo{author}{Chapin, F.S.},
  \bibinfo{author}{Ewel, J.J.}, \bibinfo{author}{Hector, A.},
  \bibinfo{author}{Inchausti, P.}, \bibinfo{author}{Lavorel, S.},
  \bibinfo{author}{Lawton, J.H.}, \bibinfo{author}{Lodge, D.M.},
  \bibinfo{author}{Loreau, M.}, \bibinfo{author}{Naeem, S.},
  \bibinfo{author}{Schmid, B.}, \bibinfo{author}{Set{\"a}l{\"a}, H.},
  \bibinfo{author}{Symstad, A.J.}, \bibinfo{author}{Vandermeer, J.},
  \bibinfo{author}{Wardle, D.A.}, \bibinfo{year}{2005}.
\newblock \bibinfo{title}{Effects of {{Biodiversity}} on {{Ecosystem
  Functioning}}: {{A Consensus}} of {{Current Knowledge}}}.
\newblock \bibinfo{journal}{Ecological Monographs} \bibinfo{volume}{75},
  \bibinfo{pages}{3--35}.
\newblock \DOIprefix\doi{10.1890/04-0922}.
%Type = Article
\bibitem[{Hu et~al.(2022)Hu, Amor, Barbier, Bunin and Gore}]{hu_emergent_2022}
\bibinfo{author}{Hu, J.}, \bibinfo{author}{Amor, D.R.},
  \bibinfo{author}{Barbier, M.}, \bibinfo{author}{Bunin, G.},
  \bibinfo{author}{Gore, J.}, \bibinfo{year}{2022}.
\newblock \bibinfo{title}{Emergent phases of ecological diversity and dynamics
  mapped in microcosms}.
\newblock \bibinfo{journal}{Science} \bibinfo{volume}{378},
  \bibinfo{pages}{85--89}.
\newblock \URLprefix \url{https://www.science.org/doi/10.1126/science.abm7841},
  \DOIprefix\doi{10.1126/science.abm7841}. \bibinfo{note}{publisher: American
  Association for the Advancement of Science}.
%Type = Article
\bibitem[{Ikegami and Hashimoto(2002)}]{ikegamiDynamicalSystemsApproach2002}
\bibinfo{author}{Ikegami, T.}, \bibinfo{author}{Hashimoto, K.},
  \bibinfo{year}{2002}.
\newblock \bibinfo{title}{Dynamical {{Systems Approach}} to {{Higher}}-level
  {{Heritability}}}.
\newblock \bibinfo{journal}{Journal of Biological Physics}
  \bibinfo{volume}{28}, \bibinfo{pages}{799--804}.
\newblock \DOIprefix\doi{10.1023/A:1021215511897}.
%Type = Article
\bibitem[{Katz et~al.(2004)Katz, Finkel, Grzebyk, Knoll and
  Falkowski}]{katzEvolutionaryTrajectoriesBiogeochemical2004}
\bibinfo{author}{Katz, M.E.}, \bibinfo{author}{Finkel, Z.V.},
  \bibinfo{author}{Grzebyk, D.}, \bibinfo{author}{Knoll, A.H.},
  \bibinfo{author}{Falkowski, P.G.}, \bibinfo{year}{2004}.
\newblock \bibinfo{title}{Evolutionary {{Trajectories}} and {{Biogeochemical
  Impacts}} of {{Marine Eukaryotic Phytoplankton}}}.
\newblock \bibinfo{journal}{Annual Review of Ecology, Evolution, and
  Systematics} \bibinfo{volume}{35}, \bibinfo{pages}{523--556}.
\newblock \DOIprefix\doi{10.1146/annurev.ecolsys.35.112202.130137}.
%Type = Article
\bibitem[{Lewontin(1970)}]{lewontinUnitsSelection1970}
\bibinfo{author}{Lewontin, R.C.}, \bibinfo{year}{1970}.
\newblock \bibinfo{title}{The {{Units}} of {{Selection}}}.
\newblock \bibinfo{journal}{Annual Review of Ecology and Systematics}
  \bibinfo{volume}{1}, \bibinfo{pages}{1--18}.
\newblock \DOIprefix\doi{10.1146/annurev.es.01.110170.000245}.
%Type = Article
\bibitem[{Liautaud et~al.(2019)Liautaud, {van Nes}, Barbier, Scheffer and
  Loreau}]{liautaudSuperorganismsLooseCollections2019}
\bibinfo{author}{Liautaud, K.}, \bibinfo{author}{{van Nes}, E.H.},
  \bibinfo{author}{Barbier, M.}, \bibinfo{author}{Scheffer, M.},
  \bibinfo{author}{Loreau, M.}, \bibinfo{year}{2019}.
\newblock \bibinfo{title}{Superorganisms or loose collections of species? a
  unifying theory of community patterns along environmental gradients}.
\newblock \bibinfo{journal}{Ecology Letters} \bibinfo{volume}{22},
  \bibinfo{pages}{1243--1252}.
\newblock \DOIprefix\doi{10.1111/ele.13289}.
%Type = Incollection
\bibitem[{Loreau(2020)}]{loreauEcosystemSuperorganismCollection2020}
\bibinfo{author}{Loreau, M.}, \bibinfo{year}{2020}.
\newblock \bibinfo{title}{The {{Ecosystem}}: Superorganism, or {{Collection}}
  of {{Individuals}}?}, in: \bibinfo{booktitle}{The {{Ecosystem}}:
  Superorganism, or {{Collection}} of {{Individuals}}?}.
  \bibinfo{publisher}{{Princeton University Press}}, pp.
  \bibinfo{pages}{218--224}.
\newblock \DOIprefix\doi{10.1515/9780691195322-019}.
%Type = Article
\bibitem[{Machado et~al.(2021)Machado, Maistrenko, Andrejev, Kim, Bork, Patil
  and Patil}]{machadoPolarizationMicrobialCommunities2021}
\bibinfo{author}{Machado, D.}, \bibinfo{author}{Maistrenko, O.M.},
  \bibinfo{author}{Andrejev, S.}, \bibinfo{author}{Kim, Y.},
  \bibinfo{author}{Bork, P.}, \bibinfo{author}{Patil, K.R.},
  \bibinfo{author}{Patil, K.R.}, \bibinfo{year}{2021}.
\newblock \bibinfo{title}{Polarization of microbial communities between
  competitive and cooperative metabolism}.
\newblock \bibinfo{journal}{Nature Ecology \& Evolution} \bibinfo{volume}{5},
  \bibinfo{pages}{195--203}.
\newblock \DOIprefix\doi{10.1038/s41559-020-01353-4}.
%Type = Article
\bibitem[{Marcus et~al.(2022)Marcus, Turner and Bunin}]{marcus_local_2022}
\bibinfo{author}{Marcus, S.}, \bibinfo{author}{Turner, A.M.},
  \bibinfo{author}{Bunin, G.}, \bibinfo{year}{2022}.
\newblock \bibinfo{title}{Local and collective transitions in
  sparsely-interacting ecological communities}.
\newblock \bibinfo{journal}{PLOS Computational Biology} \bibinfo{volume}{18},
  \bibinfo{pages}{e1010274}.
\newblock \URLprefix
  \url{https://journals.plos.org/ploscompbiol/article?id=10.1371/journal.pcbi.1010274},
  \DOIprefix\doi{10.1371/journal.pcbi.1010274}. \bibinfo{note}{publisher:
  Public Library of Science}.
%Type = Article
\bibitem[{May(1972)}]{mayWillLargeComplex1972}
\bibinfo{author}{May, R.M.}, \bibinfo{year}{1972}.
\newblock \bibinfo{title}{Will a {{Large Complex System}} be {{Stable}}?}
\newblock \bibinfo{journal}{Nature} \bibinfo{volume}{238},
  \bibinfo{pages}{413--414}.
\newblock \DOIprefix\doi{10.1038/238413a0}.
%Type = Book
\bibitem[{M{\'e}zard et~al.(1987)M{\'e}zard, Parisi and
  Virasoro}]{mezard1987spin}
\bibinfo{author}{M{\'e}zard, M.}, \bibinfo{author}{Parisi, G.},
  \bibinfo{author}{Virasoro, M.A.}, \bibinfo{year}{1987}.
\newblock \bibinfo{title}{Spin glass theory and beyond: An Introduction to the
  Replica Method and Its Applications}. volume~\bibinfo{volume}{9}.
\newblock \bibinfo{publisher}{World Scientific Publishing Company}.
%Type = Inproceedings
\bibitem[{Penn(2003)}]{pennModellingArtificialEcosystem2003}
\bibinfo{author}{Penn, A.}, \bibinfo{year}{2003}.
\newblock \bibinfo{title}{Modelling {{Artificial Ecosystem Selection}}: A
  {{Preliminary Investigation}}}, in: \bibinfo{booktitle}{Advances in
  {{Artificial Life}}}, \bibinfo{publisher}{{Springer}}. pp.
  \bibinfo{pages}{659--666}.
\newblock \DOIprefix\doi{10.1007/978-3-540-39432-7_71}.
%Type = Incollection
\bibitem[{Penn and Harvey(2004)}]{pennRoleNonGeneticChange2004}
\bibinfo{author}{Penn, A.}, \bibinfo{author}{Harvey, I.}, \bibinfo{year}{2004}.
\newblock \bibinfo{title}{The {{Role}} of {{Non-Genetic Change}} in the
  {{Heritability}}, {{Variation}}, and {{Response}} to {{Selection}} of
  {{Artificially Selected Ecosystems}}}, in: \bibinfo{booktitle}{{Artificial
  Life IX: Proceedings of the Ninth International Conference on the Simulation
  and Synthesis of Living Systems}}. \bibinfo{publisher}{The MIT Press}.
\newblock \URLprefix \url{https://doi.org/10.7551/mitpress/1429.003.0059},
  \DOIprefix\doi{10.7551/mitpress/1429.003.0059}.
%Type = Article
\bibitem[{Penney et~al.(1993)Penney, Coolen and
  Sherrington}]{penney1993coupled}
\bibinfo{author}{Penney, R.}, \bibinfo{author}{Coolen, A.},
  \bibinfo{author}{Sherrington, D.}, \bibinfo{year}{1993}.
\newblock \bibinfo{title}{Coupled dynamics of fast spins and slow interactions
  in neural networks and spin systems}.
\newblock \bibinfo{journal}{Journal of Physics A: Mathematical and General}
  \bibinfo{volume}{26}, \bibinfo{pages}{3681}.
%Type = Article
\bibitem[{Rogers et~al.(2022)Rogers, Johnson and Munch}]{rogers_chaos_2022}
\bibinfo{author}{Rogers, T.L.}, \bibinfo{author}{Johnson, B.J.},
  \bibinfo{author}{Munch, S.B.}, \bibinfo{year}{2022}.
\newblock \bibinfo{title}{Chaos is not rare in natural ecosystems}.
\newblock \bibinfo{journal}{Nature Ecology \& Evolution} \URLprefix
  \url{https://www.nature.com/articles/s41559-022-01787-y},
  \DOIprefix\doi{10.1038/s41559-022-01787-y}.
%Type = Article
\bibitem[{S{\'a}nchez et~al.(2021)S{\'a}nchez, Vila, Chang, {Diaz-Colunga},
  Estrela and {Rebolleda-Gomez}}]{sanchezDirectedEvolutionMicrobial2021}
\bibinfo{author}{S{\'a}nchez, {\'A}.}, \bibinfo{author}{Vila, J.C.C.},
  \bibinfo{author}{Chang, C.Y.}, \bibinfo{author}{{Diaz-Colunga}, J.},
  \bibinfo{author}{Estrela, S.}, \bibinfo{author}{{Rebolleda-Gomez}, M.},
  \bibinfo{year}{2021}.
\newblock \bibinfo{title}{Directed {{Evolution}} of {{Microbial Communities}}}.
\newblock \bibinfo{journal}{Annual Review of Biophysics} \bibinfo{volume}{50},
  \bibinfo{pages}{323--341}.
\newblock \DOIprefix\doi{10.1146/annurev-biophys-101220-072829}.
%Type = Article
\bibitem[{Saxe et~al.(2019)Saxe, McClelland and Ganguli}]{saxe2019mathematical}
\bibinfo{author}{Saxe, A.M.}, \bibinfo{author}{McClelland, J.L.},
  \bibinfo{author}{Ganguli, S.}, \bibinfo{year}{2019}.
\newblock \bibinfo{title}{A mathematical theory of semantic development in deep
  neural networks}.
\newblock \bibinfo{journal}{Proceedings of the National Academy of Sciences}
  \bibinfo{volume}{116}, \bibinfo{pages}{11537--11546}.
%Type = Article
\bibitem[{Schuessler et~al.(2021)Schuessler, Mastrogiuseppe, Dubreuil, Ostojic
  and Barak}]{schuesslerInterplayRandomnessStructure2021}
\bibinfo{author}{Schuessler, F.}, \bibinfo{author}{Mastrogiuseppe, F.},
  \bibinfo{author}{Dubreuil, A.}, \bibinfo{author}{Ostojic, S.},
  \bibinfo{author}{Barak, O.}, \bibinfo{year}{2021}.
\newblock \bibinfo{title}{The interplay between randomness and structure during
  learning in {{RNNs}}}.
\newblock \bibinfo{journal}{arXiv:2006.11036 [q-bio]}
  \href{http://arxiv.org/abs/2006.11036}{{\tt arXiv:2006.11036}}.
%Type = Article
\bibitem[{Sidhom and Galla(2020)}]{sidhom_ecological_2020}
\bibinfo{author}{Sidhom, L.}, \bibinfo{author}{Galla, T.},
  \bibinfo{year}{2020}.
\newblock \bibinfo{title}{Ecological communities from random generalized
  {Lotka}-{Volterra} dynamics with nonlinear feedback}.
\newblock \bibinfo{journal}{Physical Review E} \bibinfo{volume}{101},
  \bibinfo{pages}{032101}.
\newblock \URLprefix
  \url{https://link.aps.org/doi/10.1103/PhysRevE.101.032101},
  \DOIprefix\doi{10.1103/PhysRevE.101.032101}.
%Type = Misc
\bibitem[{Vessman et~al.(2023)Vessman, Guridi-Fernández, Arias-Sánchez and
  Mitri}]{vessman_novel_2023}
\bibinfo{author}{Vessman, B.}, \bibinfo{author}{Guridi-Fernández, P.},
  \bibinfo{author}{Arias-Sánchez, F.I.}, \bibinfo{author}{Mitri, S.},
  \bibinfo{year}{2023}.
\newblock \bibinfo{title}{Novel artificial selection method improves function
  of simulated microbial communities}.
\newblock \URLprefix
  \url{https://www.biorxiv.org/content/10.1101/2023.01.08.523165v1},
  \DOIprefix\doi{10.1101/2023.01.08.523165}.
%Type = Article
\bibitem[{van Vliet and Doebeli(2019)}]{vlietRoleMultilevelSelection2019}
\bibinfo{author}{van Vliet, S.}, \bibinfo{author}{Doebeli, M.},
  \bibinfo{year}{2019}.
\newblock \bibinfo{title}{The role of multilevel selection in host microbiome
  evolution}.
\newblock \bibinfo{journal}{Proceedings of the National Academy of Sciences}
  \bibinfo{volume}{116}, \bibinfo{pages}{20591--20597}.
\newblock \DOIprefix\doi{10.1073/pnas.1909790116}.
%Type = Article
\bibitem[{Williams and Lenton(2007)}]{williamsArtificialSelectionSimulated2007}
\bibinfo{author}{Williams, H.T.P.}, \bibinfo{author}{Lenton, T.M.},
  \bibinfo{year}{2007}.
\newblock \bibinfo{title}{Artificial selection of simulated microbial
  ecosystems}.
\newblock \bibinfo{journal}{Proceedings of the National Academy of Sciences}
  \bibinfo{volume}{104}, \bibinfo{pages}{8918--8923}.
\newblock \DOIprefix\doi{10.1073/pnas.0610038104}.
%Type = Article
\bibitem[{Wilson and Sober(1989)}]{wilsonRevivingSuperorganism1989}
\bibinfo{author}{Wilson, D.S.}, \bibinfo{author}{Sober, E.},
  \bibinfo{year}{1989}.
\newblock \bibinfo{title}{Reviving the superorganism}.
\newblock \bibinfo{journal}{Journal of Theoretical Biology}
  \bibinfo{volume}{136}, \bibinfo{pages}{337--356}.
\newblock \DOIprefix\doi{10.1016/s0022-5193(89)80169-9}.
%Type = Article
\bibitem[{Xie and Shou(2021)}]{xieSteeringEcologicalevolutionaryDynamics2021}
\bibinfo{author}{Xie, L.}, \bibinfo{author}{Shou, W.}, \bibinfo{year}{2021}.
\newblock \bibinfo{title}{Steering ecological-evolutionary dynamics to improve
  artificial selection of microbial communities}.
\newblock \bibinfo{journal}{Nature Communications} \bibinfo{volume}{12},
  \bibinfo{pages}{6799}.
\newblock \DOIprefix\doi{10.1038/s41467-021-26647-4}.
%Type = Article
\bibitem[{Xie et~al.(2019)Xie, Yuan and
  Shou}]{xieSimulationsRevealChallenges2019}
\bibinfo{author}{Xie, L.}, \bibinfo{author}{Yuan, A.E.}, \bibinfo{author}{Shou,
  W.}, \bibinfo{year}{2019}.
\newblock \bibinfo{title}{Simulations reveal challenges to artificial community
  selection and possible strategies for success}.
\newblock \bibinfo{journal}{PLOS Biology} \bibinfo{volume}{17},
  \bibinfo{pages}{e3000295}.
\newblock \DOIprefix\doi{10.1371/journal.pbio.3000295}.

\end{thebibliography}


\begin{thebibliography}{}

\bibitem[Barbier et~al., 2018]{barbierGenericAssemblyPatterns2018}
Barbier, M., Arnoldi, J.-F., Bunin, G., and Loreau, M. (2018).
\newblock Generic assembly patterns in complex ecological communities.
\newblock {\em Proceedings of the National Academy of Sciences},
  115(9):2156--2161.

\bibitem[Barbier et~al.,
  2021]{barbierFingerprintsHighDimensionalCoexistence2021}
Barbier, M., {de Mazancourt}, C., Loreau, M., and Bunin, G. (2021).
\newblock Fingerprints of {{High-Dimensional Coexistence}} in {{Complex
  Ecosystems}}.
\newblock {\em Physical Review X}, 11(1):011009.

\bibitem[Bunin, 2017]{buninEcologicalCommunitiesLotkaVolterra2017}
Bunin, G. (2017).
\newblock Ecological communities with {{Lotka}}-{{Volterra}} dynamics.
\newblock {\em Physical Review E}, 95(4):042414.

\bibitem[Gumbel, 2004]{gumbelStatisticsExtremes2004}
Gumbel, E.~J. (2004).
\newblock {\em Statistics of {{Extremes}}}.
\newblock {Courier Corporation}.

\bibitem[Sidhom and Galla, 2020]{sidhom_ecological_2020}
Sidhom, L. and Galla, T. (2020).
\newblock Ecological communities from random generalized {Lotka}-{Volterra}
  dynamics with nonlinear feedback.
\newblock {\em Physical Review E}, 101(3):032101.

\end{thebibliography}

\end{document}

\maketitle
\tableofcontents
\newpage
\section{Phase diagram}\label{SI:section:phase}
\begin{figure}[h]
\begin{center}
\begin{tikzpicture}
\node at (0,0) {\includegraphics[width=9.7cm]{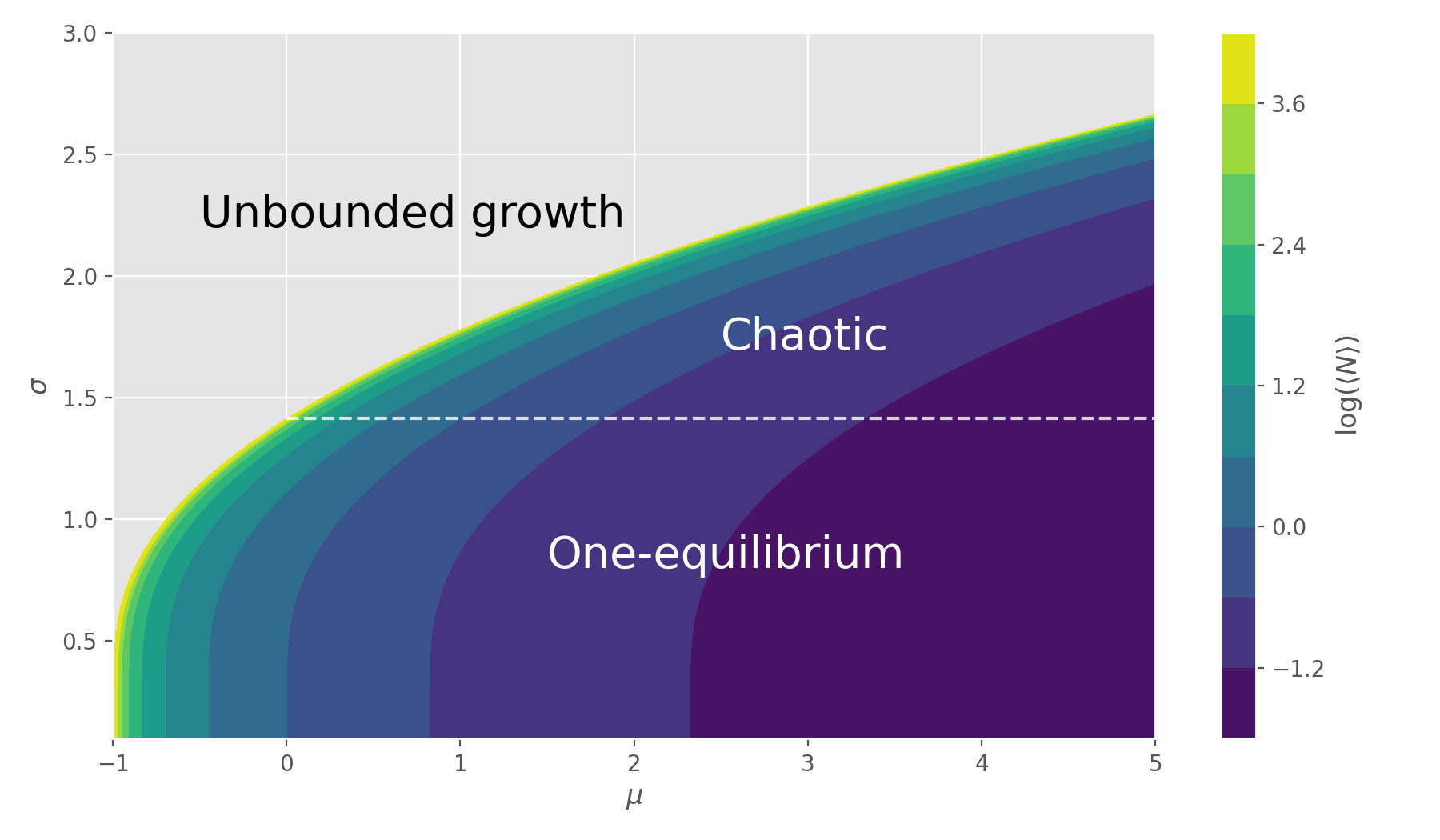}};
\end{tikzpicture}
\caption{Phase diagram of the dynamics of the Lotka-Volterra equations, superposed with the contour plot of the log of the mean population in the limit $S\rightarrow \infty$, in the space ($\mu$, $\sigma$) with $\gamma=0$ and all carrying capacities equal to one. This plot is a reproduction using the equations derived in \cite{buninEcologicalCommunitiesLotkaVolterra2017}.}
\label{SI:phase}
\end{center}
\end{figure}

The phase diagram of the random Lotka-Volterra model (equation \eqref{SI:eq:LV_eq} of the main text) in the space of the parameters $\mu$ and  $\sigma$, for the case $\gamma=0$, was derived in \cite{buninEcologicalCommunitiesLotkaVolterra2017} and is reproduced in Figure \ref{SI:phase}.
In the unbounded growth phase, some population sizes diverge in finite time: this is a pathological feature of the Lotka-Volterra equations, that can only be corrected by modifying the equations \markasnew{{\cite{sidhom_ecological_2020}}}. The chaotic phase is characterised by a chaotic dynamics with multiple unstable equilibria.
In the unique-equilibrium phase, the community converges, independently of the initial conditions, toward a unique ecological equilibrium where numerous species coexist.
In light of this result, we decided to draw our \textit{initial} interaction matrix in the unique-equilibrium phase as the ecological equilibrium is well defined.
In the last two phases, a long-term value of the mean abundance can be computed, depending only on  $\mu$, $\sigma$ and $\gamma$: the contour plot of the log of this mean value is represented in Figure \ref{SI:phase}.
Since the total abundance is proportional to the mean abundance, this surface is used in Figure 4 of the main text for the comparison between a purely random and the evolved interaction matrix.

\section{First order perturbation theory of Lotka-Volterra equations}\label{SI:first_order}
The equilibrium condition for the Lotka-Volterra equations is:

\begin{equation}\label{SI:LV-eq}
0 =N_i\left[K_i-N_i-\sum_j\alpha_{ij}N_j \right] =N_i\left[K_i-[(\mathbb{I}+\alpha)\vb{N}]_i \right]
\end{equation}
We define $\chi_{ij}=\frac{\partial N_i}{\partial K_j}$ the perturbation matrix that measure the effect of a perturbation of the carrying capacities on the equilibrium abundances. We will denote with a $^{\star}$ the vectors or matrices reduced to the set of extant species $\lbrace i=1, \dots, S \, \vert \, N_i>0\rbrace$.\\

The solution of equation \eqref{SI:LV-eq} is:
\begin{equation}\label{SI:N_eq}
\begin{aligned}
(\mathbb{I}^{\star}+\alpha^{\star}) \vb{N}^{\star}&=\vb{K}^{\star}\\
\vb{N}^{\star}&=(\mathbb{I}^{\star}+\alpha^{\star})^{-1}\vb{K}^{\star}
\end{aligned}
\end{equation}
We now differentiate equation \eqref{SI:LV-eq} with respect to $K_j$:
\begin{equation}
0 =\chi_{ij}\left[K_i-N_i-\sum_k\alpha_{ik}N_k\right]+N_i\left[\delta_{ij}-\chi_{ij}-\sum_k\alpha_{ik}\chi_{kj}\right]
\end{equation}
For the set of extant species the first term is equal to zero and $N_i\neq 0$ so we must have :
\begin{equation}
\chi_{ij}+\sum_k\alpha_{i,k}\chi_{k,j}=\delta_{ij}
\end{equation}
That is to say, in matrix notation : 
\begin{equation}
\chi^{\star}+\alpha^{\star}\chi^{\star}=\mathbb{I}^{\star}
\end{equation}
This gives us an expression for $\chi^{\star}$:
\begin{equation}\label{SI:v_mat}
\chi^{\star}=(\mathbb{I}^{\star}+\alpha^{\star})^{-1}
\end{equation}
and from equations \eqref{SI:N_eq} and \eqref{SI:v_mat} we get that at equilibrium:
\begin{equation}
\vb{N}^{\star}=\chi^{\star}\vb{K}^{\star}
\end{equation}

\section{Maximum over Gaussian samples}
\label{SI:max}

Let $\vb{x_1}, \dots, \vb{x_n}$ be independent Gaussian random vectors of dimension $d$ and of law $\mathcal{N}(0, \mathbb{I}_d)$. For $\vb{u} \in \mathbb{R}^d$, we define $f_i=\vb{x_i}\cdot \vb{u}$. We want to find the distribution of $f_{\star}=\textrm{max}(\lbrace f_i\rbrace)$ and of the associated $\vb{x_{\star}}$. \\

We denote $\vu{u}=\frac{\vb{u}}{\|\vb{u}\|}$. Let $P_u$ and $P_{u^{\bot}}$ be the projection matrices on $\vu{u}$ and on its orthogonal space $\vu{u}^{\bot}$ such that we have $f_i = \|\vb{u}\|P_u \vb{x_i}$. By Cochran theorem $P_u \vb{x_i}$ and $P_{u^{\bot}} \vb{x_i}$ are independent Gaussian variables of law $\mathcal{N}(0, P_u)$ and $\mathcal{N}(0, P_{u^{\bot}})$. \\

\begin{figure}
\begin{center}
\includegraphics[width=8cm]{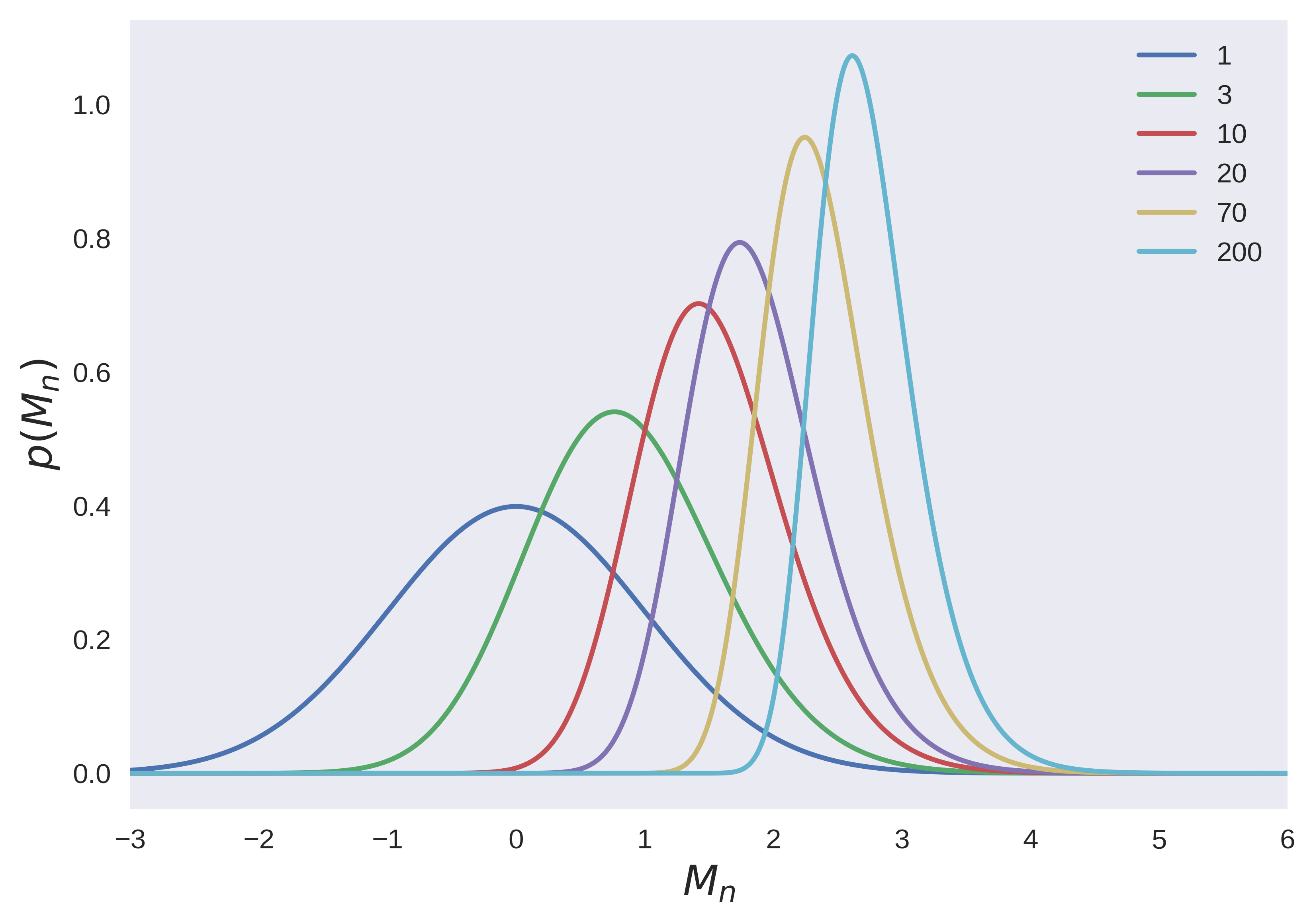}
\includegraphics[width=8cm]{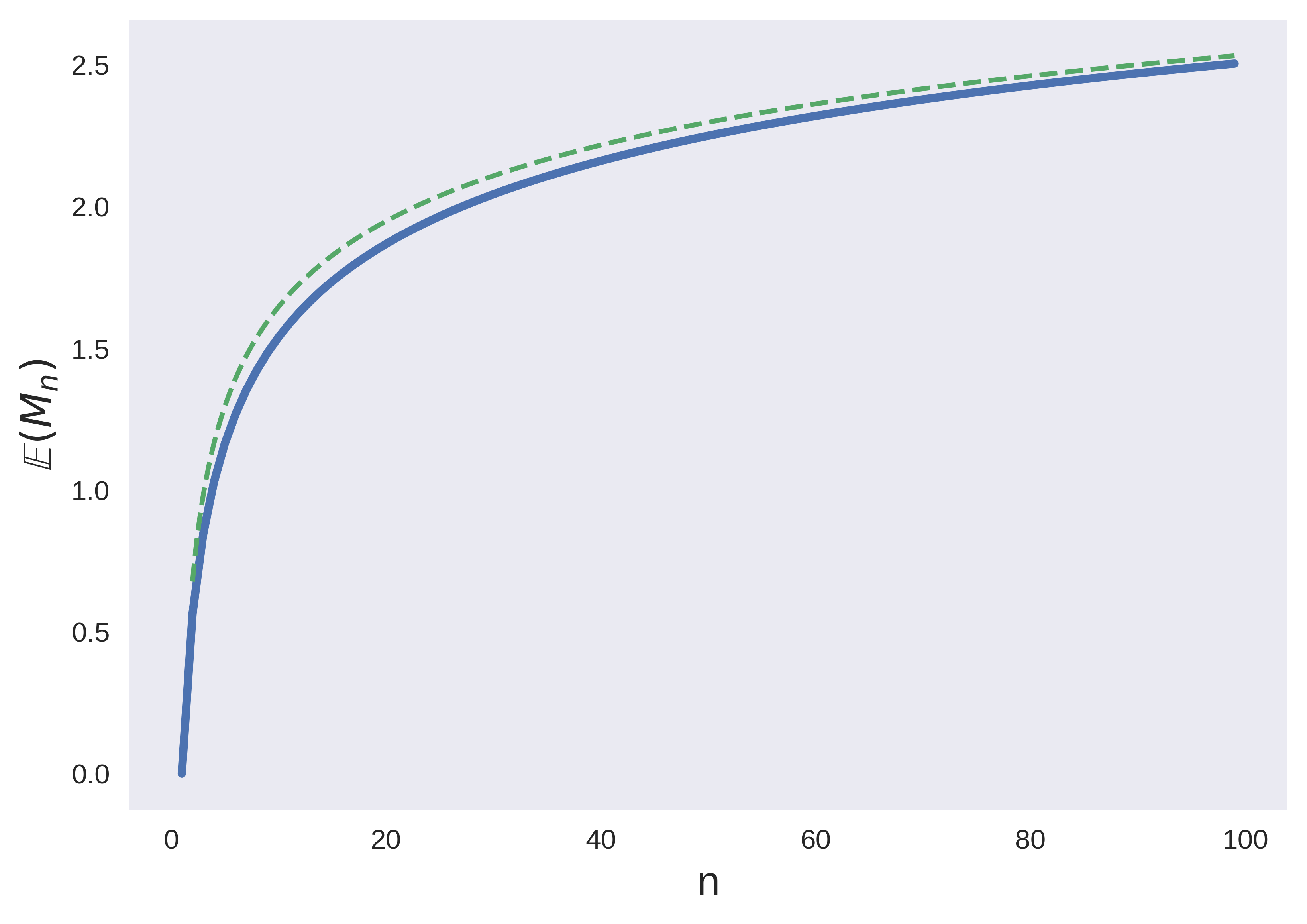}
\caption{At the top is the distribution of $M_n$ for different values of n. At the bottom is the evolution of the expected value of this distribution with $n$ (in blue), with an approximation by $0.5-\sqrt{2\log(n)}$ in dotted green.}
\label{SI:distrib}
\end{center}
\end{figure}

As $P_u \vb{x_i}$ is aligned with $\hat u$ and because $\hat u$ is normalized, we have $P_u\vb{x_i}=y_i\vu{u}$ with $y_i \, \sim \, \mathcal{N}(0,1)$. We thus have $f_i=\|\vb{u}\|y_i$. In this notation $f_i>f_j \Leftrightarrow y_i>y_j$.
We define $M_n =\mathrm{max}(y_1, \dots, y_n)$ such that  $y_{\star}=M_n$ if $\, \forall i\neq \star  \, , \, y_i<y_{\star}$ . By Bayes formula, $M_n$ has the probability density :
\begin{equation}
\begin{aligned}
p(M_n=y) &= p(y_{\star}=y\, | \, \forall i\neq \star  \, , \, y_i<y_{\star}) \\ &= \frac{1}{Z}\mathbb{P}(\forall i\neq \star  \, , \, y_i<y)p(y_{\star}=y) \\
&= \frac{1}{Z}\left(\mathbb{P}(y_i<y)\right)^{n-1}p(y_{\star}=y) \\
&= \frac{1}{Z} \Phi(y)^{n-1}\phi(y)
\end{aligned}
\end{equation}
with $\Phi$ and $\phi$ the CDF and PDF of Gaussian law and $Z$ a normalisation constant :
\begin{equation}
Z = \int_{-\infty}^{+\infty}\Phi(y)^{n-1}\phi(y)\, \textrm{d}y = \frac{1}{n}\left[\Phi(y)^n\right]_{-\infty}^{+\infty}=\frac{1}{n}
\end{equation}
Finally, $p(M_n=y)=n\Phi(y)^{n-1}\phi(y)$. This distribution is represented in Figure \ref{SI:distrib}. It is asymptotic to a shifted Gumbel distribution \cite{gumbelStatisticsExtremes2004}.\\

We now use the decomposition $\vb{x_{\star}}=P_u \vb{x_{\star}}+P_{u^{\bot}} \vb{x_{\star}} = M_n\vu{u}+P_{u^{\bot}} \vb{x_{\star}}$.
We thus have :
\begin{equation}
\vb{x_{\star}} =  M_n\vu{u}+\vb{b} \quad \text{with} \quad \vb{b} \, \sim \, \mathcal{N}(0, P_{u^{\bot}})
\end{equation} or, in another notation :
\begin{equation}\label{eq:decomposition}
\vb{x_{\star}} =   M_n\vu{u}+P_{u^{\bot}}\vb{z} = (M_n-\vb{z}\cdot \vu{u})\vu{u}+\vb{z} \quad \text{with} \quad \vb{z}\, \sim \, \mathcal{N}(0, \mathbb{I}_d)
\end{equation}

\section{Mutation-Selection Process for any $\gamma$ }\label{SI:section:mut_sel}

Let $\vb{N}$ be the equilibrium for the community described by the Lotka-Volterra equations. Each species' abundance obeys the equation:
\begin{equation}\label{SI:eq:LV_eq}
0 = N_i\left[K_i-N_i-\sum_{j}\alpha_{ij}N_j \right].
\end{equation}

After a mutational step, the interaction matrix becomes, at first order in $\varepsilon$ :
\begin{equation}\label{SI:eq:alpha_hat}
\hat \alpha_{ij}=\alpha_{ij}+\frac{\varepsilon\sigma}{\sqrt{S}}\eta_{ij}
\end{equation} with $\eta$ a Gaussian matrix of expected value zero, variance $1$ and symmetric correlation $\gamma$. \\

Let $\vu{N}$ be the equilibrium abundances associated to the interaction matrix $\hat \alpha$. To first order in $\varepsilon$, the equation for $\vu{N}$ is:

\begin{equation}
0= \hat N_i\left[K_i-\hat N_i- \sum_{j}\alpha_{ij}\hat N_j \underbrace{-\varepsilon\frac{\sigma}{\sqrt{S}} \sum_{j}\eta_{ij}N_j}_{\delta K_i} \right]
\end{equation}
The variable $\vu{N}$ is the solution of equation \eqref{SI:eq:LV_eq}, where we added a perturbation field $\delta\vb{K}=-\varepsilon\frac{\sigma}{\sqrt{S}}\eta \vb{N}$ of order $\varepsilon$. 
Defining $\chi_{ij}=\frac{\partial N_i}{\partial K_j}$ as in section \ref{SI:first_order},  for small $\varepsilon$
we can compute the resulting change in abundances $\delta \vb{N}$ as:
\begin{equation}
\begin{aligned}
\delta \vb{N} &=\chi\delta\vb{K} \\
&=-\varepsilon\frac{\sigma}{\sqrt{S}} \chi\eta \vb{N}
\end{aligned}
\end{equation}

\markasnew{This change in abundance $\delta \vb{N}$ will result in a change in the community function $f(\vb{N})$ that we can characterise:}
\begin{equation}
\begin{aligned}
\delta f &=\grad f(\vb{N})\cdot\delta \vb{N}\\
&=-\varepsilon\frac{\sigma}{\sqrt{S}} \grad f(\vb{N})^{\top}\chi\eta \vb{N}\\
&=-\varepsilon\frac{\sigma}{\sqrt{S}} [(\chi^{\top}\grad f(\vb{N})) \otimes \vb{N}]:\eta \\
&=-\varepsilon\frac{\sigma}{\sqrt{S}} [\vb{v} \otimes \vb{N}]:\eta
\end{aligned}
\end{equation}
\markasnew{with $\vb{v} =\chi^{\top}\grad f(\vb{N})$,} $\otimes$ the tensor product ($(u\otimes v)_{ij}=u_iv_j$) and $:$ the tensor contraction ($A:B = \sum_{ij}A_{ij}B_{ij}$). Because in the case $\gamma \neq0$, the $\eta_{ij}$ are not all independent, we can't directly apply the results of section {\ref{SI:max}}. For this reason, we use the decomposition of $\eta$ (see SI of \cite{barbierGenericAssemblyPatterns2018}) :

\begin{equation}\label{SI:eq:sym_mat}
    \eta = \frac{x+\kappa x^{\top}}{\sqrt{1+\kappa^2}}
\end{equation}
with $\kappa=\frac{1-\sqrt{1-\gamma^2}}{\gamma}$ and $x$ a Gaussian matrix of mean 0, variance 1 with no correlations between $x_{ij}$ and $x_{ji}$ for $i\neq j$.

Using $A:x^{\top}=A^{\top}:x$, we have 
\begin{equation}
   \delta f = -\varepsilon\frac{\sigma}{\sqrt{S}\sqrt{1+\kappa^2}}\left[\vb{v} \otimes \vb{N}+\kappa  \vb{N}\otimes\vb{v}\right]: x
\end{equation}

\markasnew{We now denote $u=-\varepsilon\frac{\sigma}{\sqrt{S}\sqrt{1+\kappa^2}}\left[\vb{v} \otimes \vb{N}+\kappa  \vb{N}\otimes\vb{v}\right] $ so that $\delta f = u:x$ as in section {\ref{SI:max}}, but this time $x$ and $u$ are matrices instead of vectors and the scalar product is replaced by a tensor contraction. However, by packing the two indices $ij$ of $x_{ij}$ and $u_{ij}$ into a general index $\alpha$ we can interpret the exact same form as the scalar product of two large vectors of dimension $d=S^2$. In this way, the tensor contraction reduces to a simple scalar product so that we can directly apply the results of section {\ref{SI:max}}. 

Among $n$ different realisations of the random matrix $x$ (and thus of $\eta$), the one that will give rise to the highest score can be written:}
\begin{equation}
    x_{\star}=M_n\frac{\vb{u}}{\|\vb{u}\|}=\frac{-M_n}{\mathcal{N}}\left[\vb{v} \otimes \vb{N}+\kappa  \vb{N}\otimes\vb{v}\right]+B
\end{equation}
\markasnew{and is associated with a change in score:}
\begin{equation}
    \delta f_\star=u:x_\star =\frac{\varepsilon\sigma\mathcal{N}}{\sqrt{S}\sqrt{1+\kappa^2}}M_n
\end{equation}
with $\mathcal{N}=\| \vb{v} \otimes \vb{N}+\kappa  \vb{N}\otimes\vb{v}\|$, \markasnew{$M_n$ the random variable defined in section} \ref{SI:max} and $B$ a Gaussian matrix "orthogonal to the selected direction" \markasnew{($u:B=0$)}. This equation reduces to equation (9) of the main text when $\gamma=0$.

Putting the expression of $x_\star$ in the formula for $\eta$ and using $\frac{2\kappa}{1+\kappa^2}=\gamma$ we get an expression for the selected $\eta_\star$:
\begin{equation}\label{SI:selection}
\begin{aligned}
\eta_\star&=\frac{-M_n}{\mathcal{N}\sqrt{1+\kappa^2}}\left[(1+\kappa^2) \vb{v} \otimes \vb{N}+2\kappa  \vb{N}\otimes \vb{v}\right]+\Tilde{B}\\
&=\frac{-M_n\sqrt{1+\kappa^2}}{\mathcal{N}}\left[ \vb{v} \otimes \vb{N}+\gamma \vb{N}\otimes \vb{v}\right]+\Tilde{B}\\
\end{aligned}
\end{equation} with $\Tilde{B} = \frac{B+\kappa B^{\top}}{\sqrt{1+\kappa^2}}$ being a Gaussian matrix of mean 0 and variance 1 with symmetric correlation $\gamma$ (it is the same construction as equation \eqref{SI:eq:sym_mat}).

\markasnew{The resulting selected interaction matrix can be written:}
\begin{equation}\label{SI:eq:alpha_rec}
    \hat \alpha_\star=\alpha-\frac{\varepsilon\sigma}{\sqrt{S}}\left( \frac{M_n\sqrt{1+\kappa^2}}{\mathcal{N}}\left[ \vb{v} \otimes \vb{N}+\gamma \vb{N}\otimes \vb{v}\right]-\Tilde{B}\right)
\end{equation}

In the case $\gamma=0$, we obtain equation (12) of the main text.

\section{Correlation of the eigenvector with the abundances at equilibrium}
\begin{figure}[h]
    \centering
    \includegraphics[width=10cm]{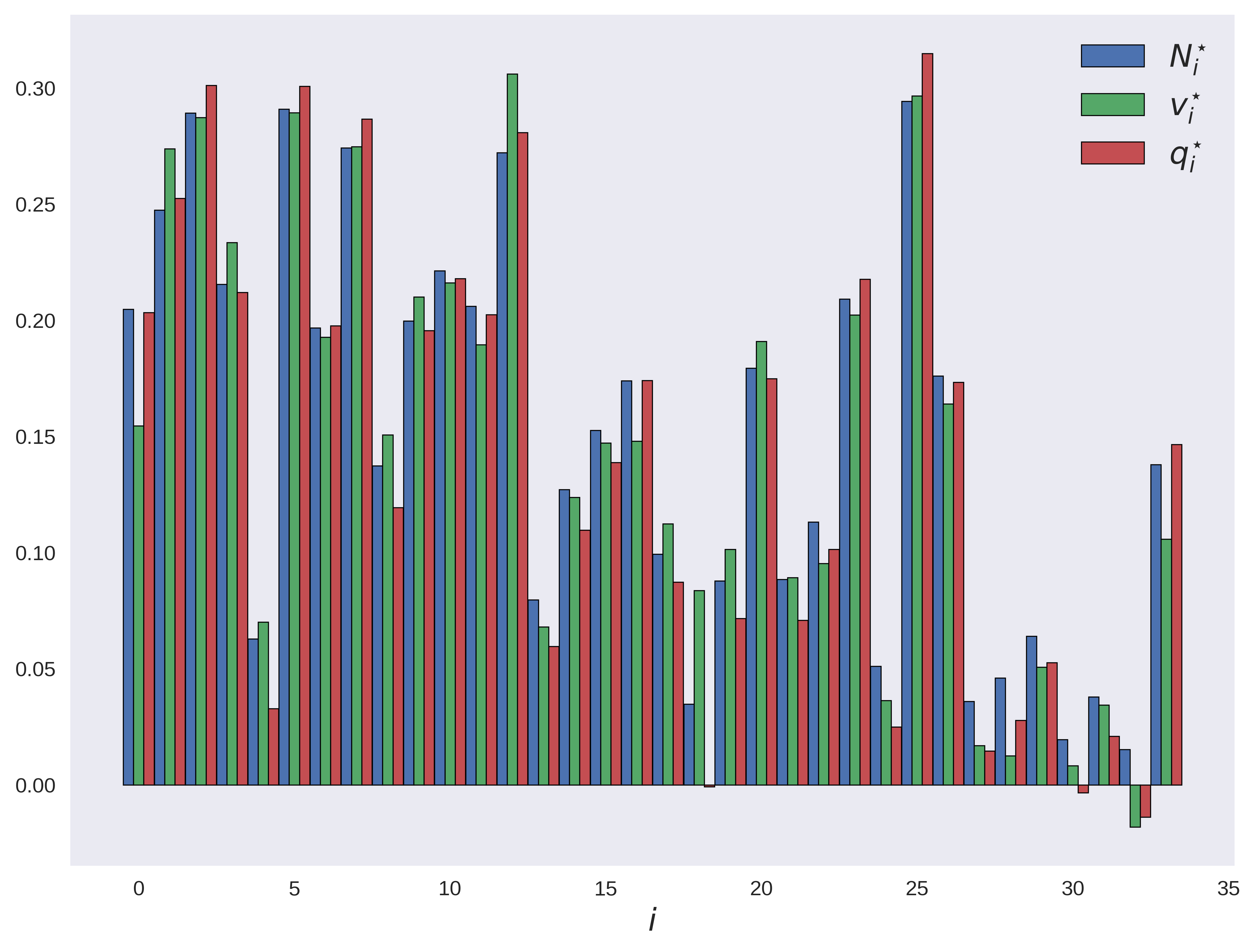}
    \caption{Coefficients of $\vb{N^\star}$, $\vb{v^\star}$ (as defined in the main text) and of the left eigenvector $\vb{q^\star}$ associated to the isolated eigenvalue of $\alpha^\star$ at generation $\tau=2000$ reduced to extant species, all normalised, with indices sorted by decreasing carrying capacity. It is evident that $\vb{N^\star}$, $\vb{v^\star}$ and $\vb{q^\star}$ are strongly correlated. Their dependence on $\vb{K}$ is weaker, but still sufficient for the structure to emerge in Fig. 5 of the main text.}
    \label{SI:fig:coeffs_Nv}
\end{figure}

\begin{figure}[h]
    \centering
    \includegraphics[width=10cm]{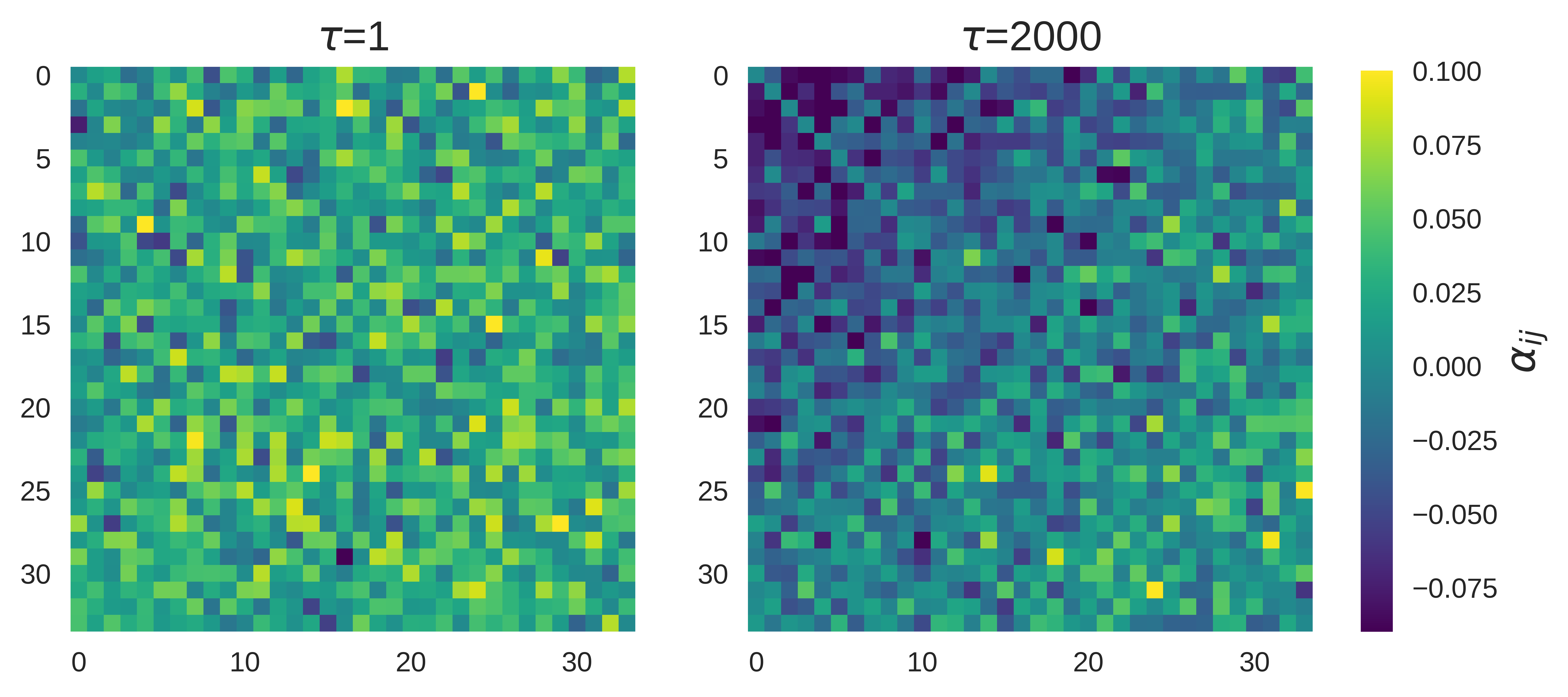}
    \caption{Coefficients of the interaction matrix $\alpha$ at generations 1 (left) and 2000 (right) with rows and columns sorted by decreasing \textbf{equilibrium abundances at $\tau=2000$} for the same simulation as described in Materials and Methods. Only the species that have positive abundance at generation $2000$ are shown.}
    \label{SI:fig:coeff_alpha_Ns}
\end{figure}

 Fig. \ref{SI:fig:coeffs_Nv} shows that at a late generation ($\tau=2000$), the vector of abundances at equilibrium is strongly correlated with the eigenvector associated to the left-most isolated eigenvalue of $\alpha$ and to the vector $\vb{v}=\frac{\delta N_T}{\delta \vb{K}}$. Theses correlations explain the structure of $\alpha$: Fig. \ref{SI:fig:coeff_alpha_Ns} shows that the species that are the most abundant are also the most mutualist.

This correlation is caused by two effects that feed back onto one another. First, species with mutualistic interactions are more likely to be abundant than those with competitive interactions. Secondly, as shown in Eq. 7 (Main text), the most abundant species will see their interactions evolve faster toward mutualism.

\section{Limit of small $\sigma$ and large $S$}\label{SI:section:small_sig}
For simplicity, we consider here that all the carrying capacities are equal (we set $K_i=1$, without loss of generality). When $S \gg 1$, $\sigma (\tau)$ is almost constant, and we assume here that it is initially small.

For small $\sigma$, the interaction matrix $\alpha(\tau)$ can be characterised only by its mean value : 
\begin{equation}
    \alpha_{ij}(\tau)=\frac{\mu(\tau)}{S}+\mathcal{O}(\sigma)
\end{equation} that we write in matrix notation :
\begin{equation}
    \alpha(\tau)=\frac{\mu(\tau)}{S}\vb{1}\vb{1}^{\top}+\mathcal{O}(\sigma)
\end{equation}
Then, following eq \eqref{SI:v_mat}, the perturbation matrix can be expressed as :
\begin{equation}
\begin{aligned}
    \chi(\tau)&=(\mathbb{I}+\alpha(\tau))^{-1} \\
        &=\left(\mathbb{I}+\frac{\mu(\tau)}{S}\vb{1}\vb{1}^{\top}+\mathcal{O}(\sigma)\right)^{-1}\\
        &=\mathbb{I}-\frac{\mu(\tau)/S}{1+\mu(\tau)}\vb{1}\vb{1}^{\top}+\mathcal{O}(\sigma)
\end{aligned}
\end{equation}
using Sherman–Morrison formula.
With this, we can compute the abundances at equilibrium:
\begin{equation}
    \begin{aligned}
    \vb{N}(\tau)&=\chi(\tau)\vb{1} \\
    &=\frac{1}{1+\mu(\tau)}\vb{1}+\mathcal{O}(\sigma)
    \end{aligned}
\end{equation}

The total abundance along the evolutionary trajectory depends, as well as the interaction matrix, only on the mean interaction strength:
\begin{equation}\label{SI:eq:dNdt}
\frac{N_T(\tau)}{S} = \langle N_i(\tau)\rangle=\frac{1}{1+\mu(\tau)}+\mathcal{O}(\sigma).
\end{equation}

In the same fashion, we get that :
\begin{equation}
    \vb{v}(\tau)=\frac{1}{1+\mu(\tau)}\vb{1}+\mathcal{O}(\sigma)
\end{equation}

Equation \eqref{SI:eq:alpha_rec} for $\gamma=0$ or equation (6) of the main text then gives the recursive equation for $\mu(\tau)$:

\begin{equation}
    \mu(\tau+1)=\mu(\tau)-\sqrt{1+\gamma} \, \frac{\varepsilon M_n\sigma}{\sqrt{S}}
\end{equation}

that has for solution:
\begin{equation}\label{SI:eq:dmudt}
\mu(\tau) = \mu(0)-\tau\,\frac{\varepsilon\sigma}{\sqrt{S}} \overline{M_n}\sqrt{1+\gamma}  
\end{equation} 

\begin{figure}[h]
    \centering
    \includegraphics[width=9cm]{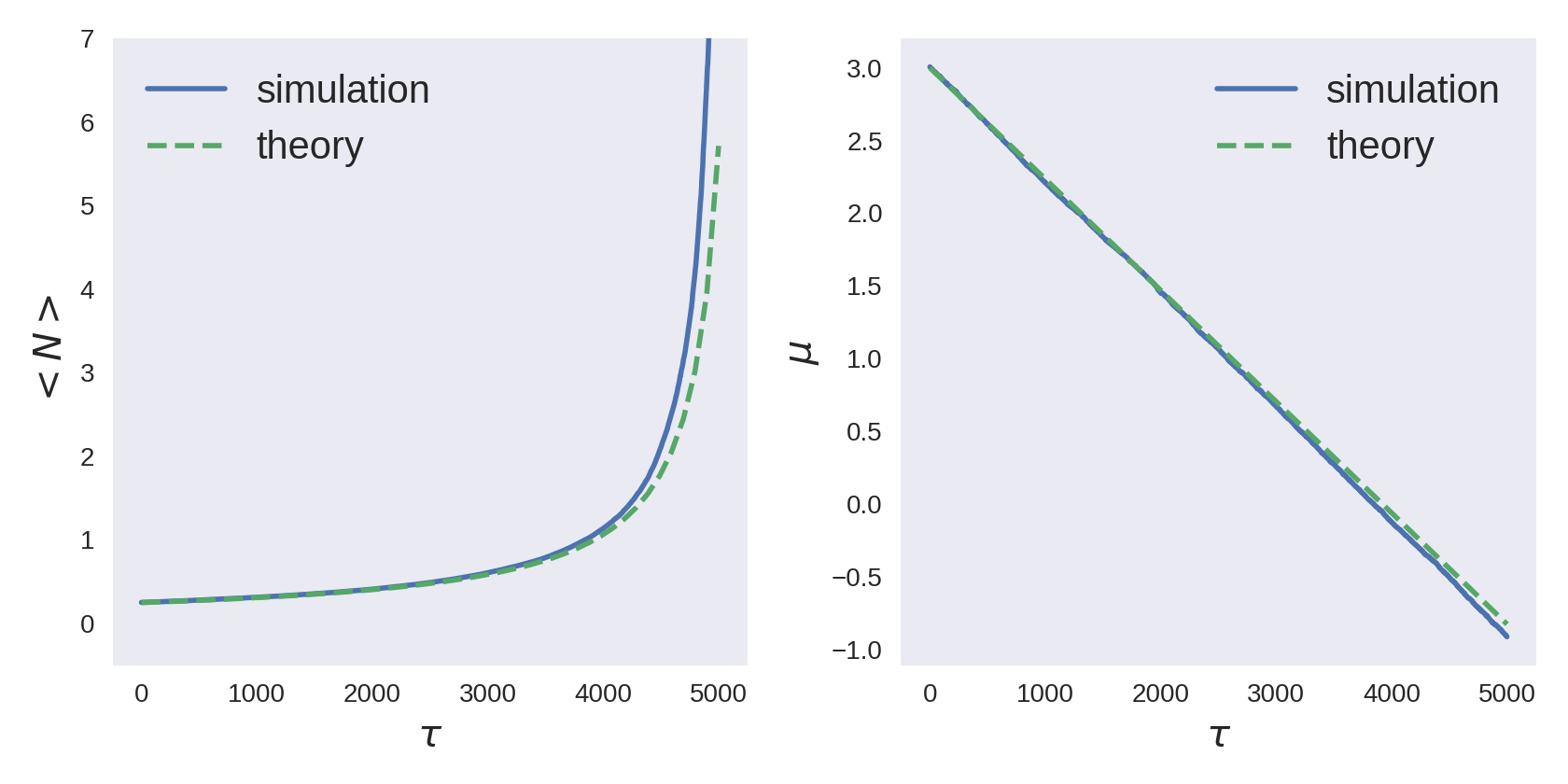}
    \caption{Comparison of the evolution of the mean abundance $<N>$ and of the rescaled mean interaction $\mu$ from the Eqs. \eqref{SI:eq:dmudt} and \eqref{SI:eq:dNdt} (green lines) and from numerical simulations (blue lines) with $S=100$ and $\sigma(0)=0.1$.}
    \label{SI:fig:small_sig}
\end{figure}

Fig. \ref{SI:fig:small_sig} shows that equations \eqref{SI:eq:dNdt} and \eqref{SI:eq:dmudt} match the numerical simulations remarkably well in the case of an initial $\sigma=0.1$ and $S=100$.

\newpage 

\section{Community changes in the absence of selection}\label{SI:random_walk}
We now look at the evolutionary dynamics of the interaction matrix in the neutral regime, when any community leaves an offspring with equal probability. 

In this subsection only, we denote for simplicity $d=S(S-1)$ the number of interaction terms and every sum $\sum $ are index on $i\neq j$. The notations $b_{ij}$ is not the same as before.

The process is the following:
\begin{itemize}
\item We have $\alpha_{ij}^{(t)}=m_t+\mathrm{std}_t a_{ij}^{(t)}$ with $m_t = \frac{1}{d}\sum\alpha_{ij}^{(t)}$ the empirical mean and $\mathrm{std}_t = \sqrt{\frac{1}{d}\sum (\alpha_{ij}^{(t)}-m_t)^2}$ the empirical standard deviation. 
\item We then define $\alpha_{ij}^{(t+1)}=m_t+\mathrm{std}_t b_{ij}^{(t+1)}$ with $b_{ij}^{(t+1)}=\frac{a_{ij}^{(t)}+\varepsilon\eta_{ij}^{(t+1)}}{\sqrt{1+\varepsilon^2}}$.
\item We repeat the operation.
\end{itemize}

Note that by definition, $a_{ij}$ have the following properties: 
$\sum a_{ij}^{(t)}=0$ and $\frac{1}{d}\sum (a_{ij}^{(t)})^2 =1$.

We now wish to get a an expression of $\mu_{t+1}$ as a function of $\mu_t$ and $\eta$. We start with the empirical mean:
\begin{equation}
\begin{aligned}
m_{t+1} &=\frac{1}{d}\sum\alpha_{ij}^{(t+1)} \\
&=m_t+\frac{\mathrm{std}_t}{d}\sum b_{ij}^{(t+1)} \\
&=m_t+\frac{\mathrm{std}_t \, \varepsilon }{\sqrt{1+\varepsilon^2}}\frac{1}{d}\sum \eta_{ij}^{(t+1)}
\end{aligned}
\end{equation}
Using the Central Limit Theorem and writing with $\mu$, $S$ and $\sigma$ we get:

\begin{equation}
\boxed{
\mu_{t+1} \sim \mathcal{N}(\mu_t, \sigma_t \frac{\varepsilon}{\sqrt{S(1+\varepsilon^2)}}) }
\end{equation}

To have a similar expression for $\sigma$ the computation is a bit longer. We start with the empirical variance:
\begin{equation*}
\begin{aligned}
\mathrm{Var}_{t+1} &= \frac{1}{d}\sum (\alpha_{ij}^{(t+1)} - m_{t+1})^2 \\
&= \frac{1}{d}\sum(m_t-m_{t+1}+\mathrm{std}_t b_{ij}^{(t+1)} )^2 \\
&= \frac{\mathrm{Var}_t}{d}\sum \left( \frac{a_{ij}^{(t)}+\varepsilon\eta_{ij}^{(t+1)}}{\sqrt{1+\varepsilon^2}} - \frac{\mathrm{std}_t \, \varepsilon }{\sqrt{1+\varepsilon^2}}\frac{1}{N}\sum_{k\neq l} \eta_{k,l}^{(t+1)}\right)
\end{aligned}
\end{equation*}

We denote $\mean{\eta}=\frac{1}{d}\sum_{k\neq l} \eta_{k,l}^{(t+1)}$ the empirical mean of $\eta$ so that:
\begin{equation*}
\begin{aligned}
\mathrm{Var}_{t+1} &= \frac{\mathrm{Var}_t}{(1+\varepsilon)^2 d}\sum \left(\alpha_{ij}^t +\varepsilon (\eta_{ij}-\mean{\eta})\right) \\
&= \frac{\mathrm{Var}_t}{(1+\varepsilon)^2 d} \left[ \sum (a_{ij}^t)^2 + \varepsilon \sum (\eta_{ij} - \mean{\eta} )^2 + 2\varepsilon \sum a_{ij}^{(t)} (\eta_{ij} -\mean{\eta}) \right]
\end{aligned}
\end{equation*}

We now use the fact that $\sum a_{ij}^{(t)}=0$ and $\frac{1}{d}\sum (a_{ij}^{(t)})^2 =1$. Furthermore, as $\eta$ is a Gaussian random variable, $\sum (\eta_{ij} - \mean{\eta} )^2 \sim \chi_{n-1}^2 $ (chi-square distribution).

As $a_{ij}$ and $\eta_{ij}$ are uncorrelated we have by the Central Limit Theorem:
\begin{equation}
\frac{1}{d}\sum a_{ij} \eta_{ij} \sim \mathcal{N}(0,\frac{1}{\sqrt{N}})
\end{equation}

We then use the asymptotic expression: $\chi_{d-1}^2 \longrightarrow d-1+\sqrt{2(d-1)} \, \mathcal{N}(0,1)$ so that in the large $S$ (and thus $d$) limit:
\begin{equation*}
\mathrm{Var}_{t+1} = \frac{\mathrm{Var}_t}{(1+\varepsilon)^2}\left[1+\varepsilon^2 + \varepsilon^2 \sqrt{\frac{2}{d}}\mathcal{N}(0,1) + \frac{2 \varepsilon}{\sqrt{d}}\mathcal{N}(0,1) \right]
\end{equation*}
But as $\varepsilon \ll 1$ we can neglect the first normal distribution:
\begin{equation}
\mathrm{Var}_{t+1} = \mathrm{Var}_t\left(1+\frac{2 \varepsilon}{1+\varepsilon^2}\frac{1}{\sqrt{d}}\mathcal{N}(0,1) \right)
\end{equation}
Using the Taylor expansion of the square root we get the expression for $\sigma$:
\begin{equation}
\boxed{
\sigma_{t+1} \sim \mathcal{N}(\sigma_t, \sigma_t \frac{\varepsilon}{S(1+\varepsilon^2)}) }
\end{equation}

\begin{figure}[h]
    \centering
    \includegraphics[scale = 0.36]{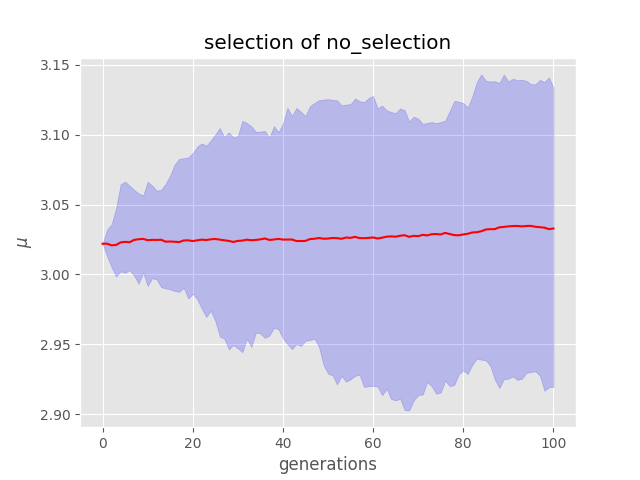}
    \includegraphics[scale = 0.36]{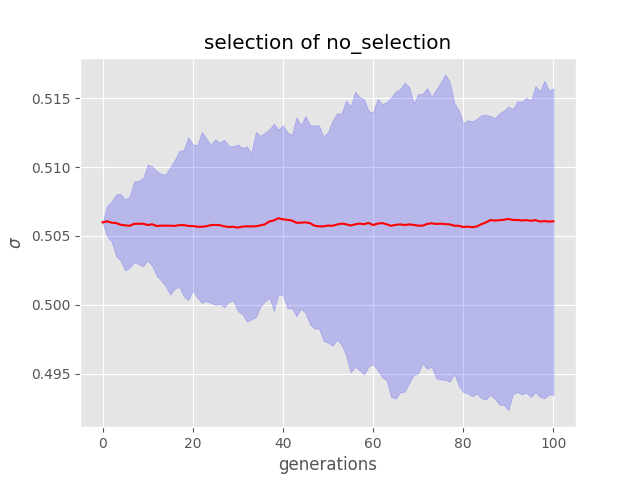}
    \includegraphics[scale = 0.36]{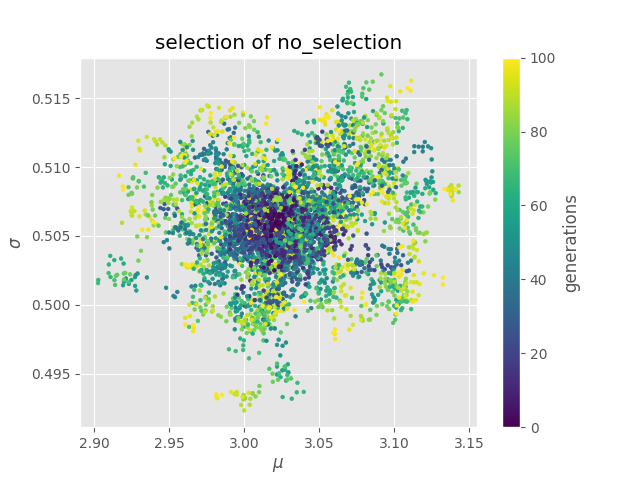}
    \includegraphics[scale = 0.36]{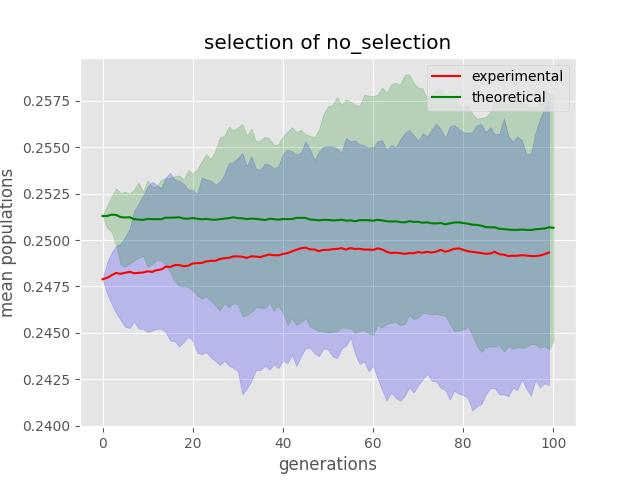}
    \caption{Evolution of the empirical $\mu$ (top-left), $\sigma$ (top-right) and mean population $\mean{N}$ (bottom-right) in absence of selection, with the parameters $S=100$, $\varepsilon=0.1$, $\tau_{\mathrm{max}}=100$, $n=50$ and $\gamma=0$. For the panels in the top and in the bottom-right, the red lines are the mean of the quantity of interest and the blue area is bounded by the maximum and minimum values. The green line in the bottom-right panel represent the predicted value from the cavity method. The panel in the bottom-left is the evolution of the communities in the $(\mu, \sigma)$ space.}
    \label{SI:fig:no_selection}
\end{figure}

We thus observe that this process is a isotropic random walk in the space $(\mu, \sigma)$ and that the variation in $\sigma$ are small compared to the one in $\mu$: it scales like $\frac{1}{S}$ for $\sigma$ and  $\frac{1}{\sqrt{S}}$ for $\mu$. 
Thus, for large $S$, $\sigma$ evolves with a longer time-scale than $\mu$ and so can be approximated as constant.
These results are consistent with the numerical observation in Figure \ref{SI:fig:no_selection}.

\section{Evolution of species diversity}

\begin{figure}
    \centering
    \includegraphics[width=8cm]{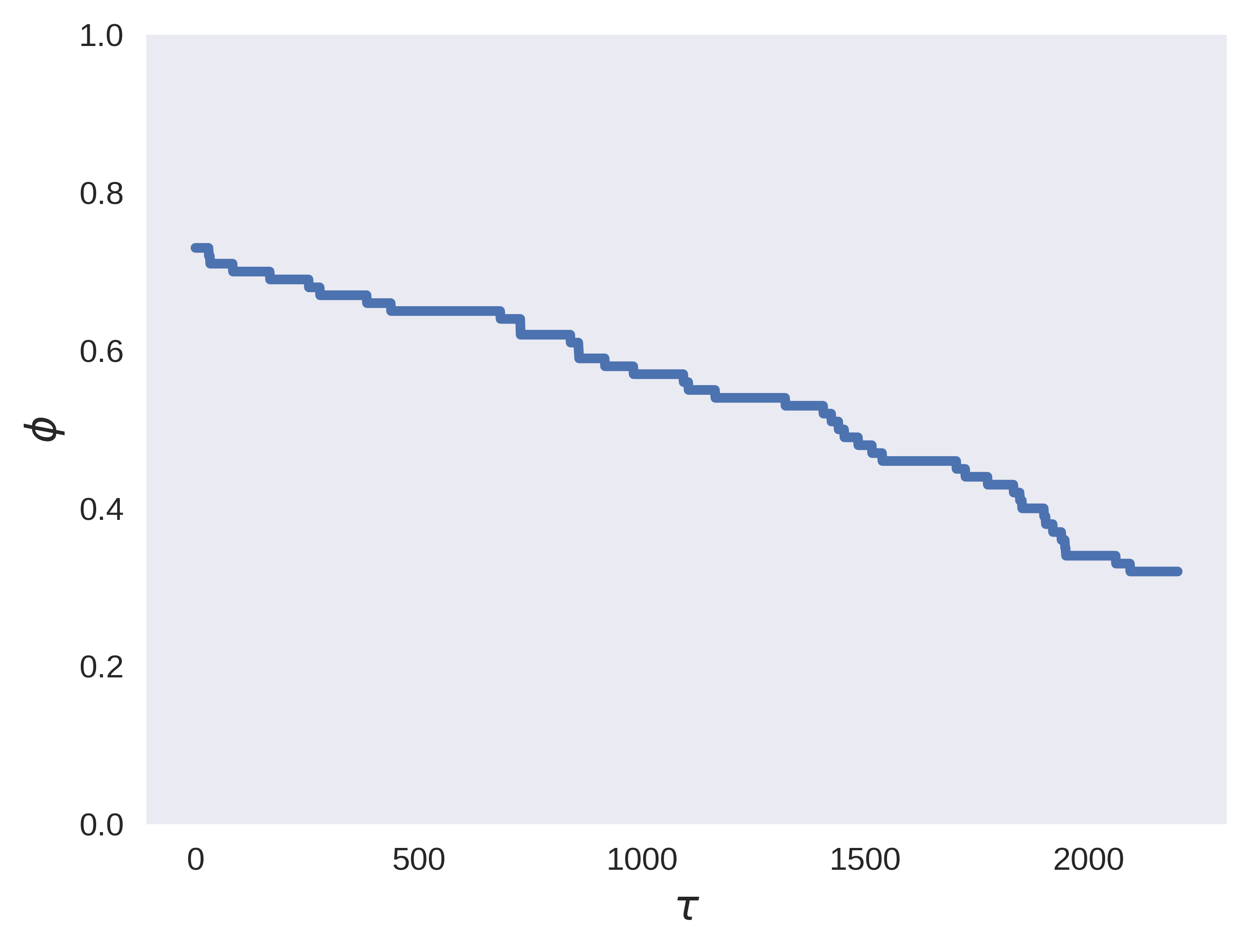}
    \caption{{\bf Changes of species richness along an evolutionary trajectory.} Selection for increased total abundance leads to a decrease in the species richness. The parameters of the simulation are described in the Methods section.}
    \label{fig:SI:diversity}
\end{figure}

Fig. \ref{fig:SI:diversity} represents the evolution of the diversity (or species richness) $\phi=\frac{S^\star}{S}$ along the same evolutionary trajectory as the figures of the main text. This shows that diversity decreases along the evolutionary trajectory but the number of extant species does not collapse, and the community maintains a substantial amount of diversity. 

\section{Comparison with synthetic interaction matrix}\label{SI:section:max_ent}
\begin{figure}
    \centering
    \includegraphics[width=8cm]{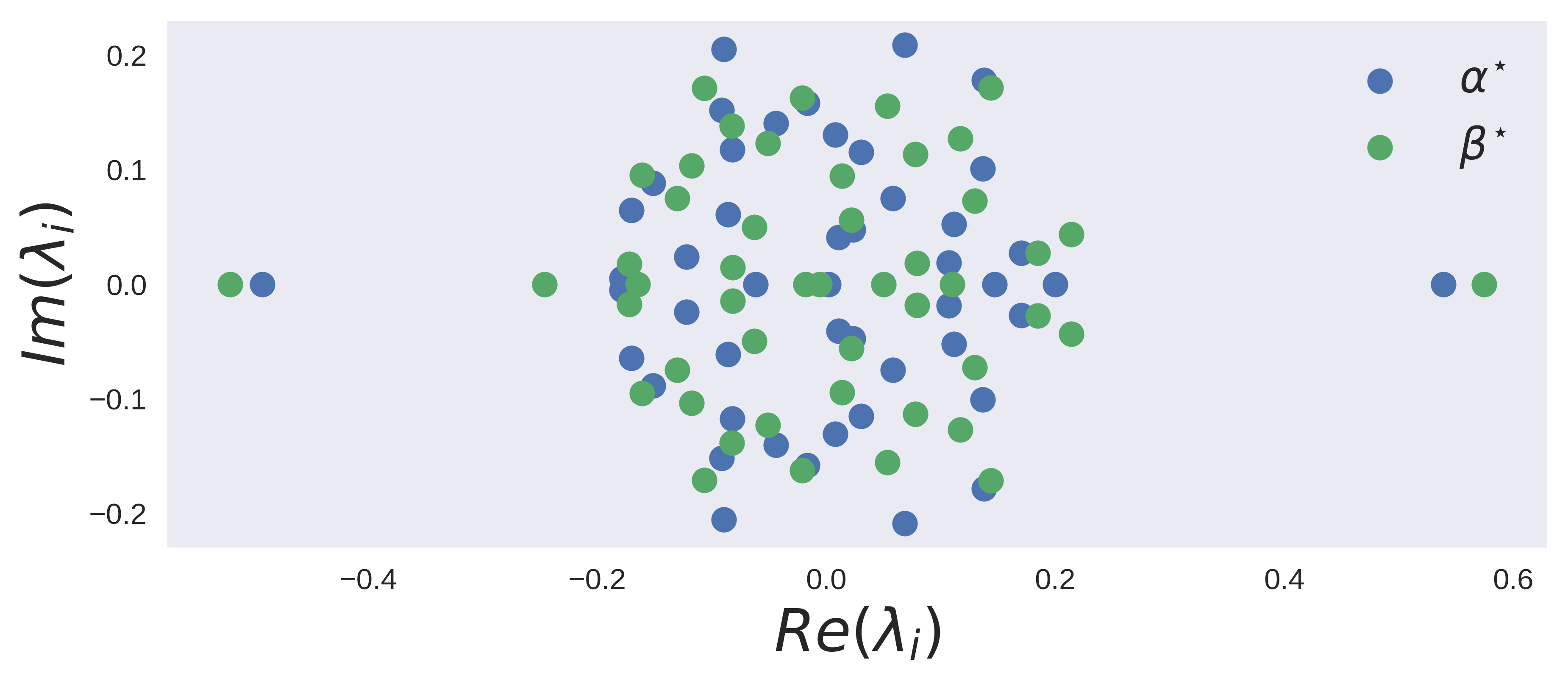}
    \includegraphics[width=8cm]{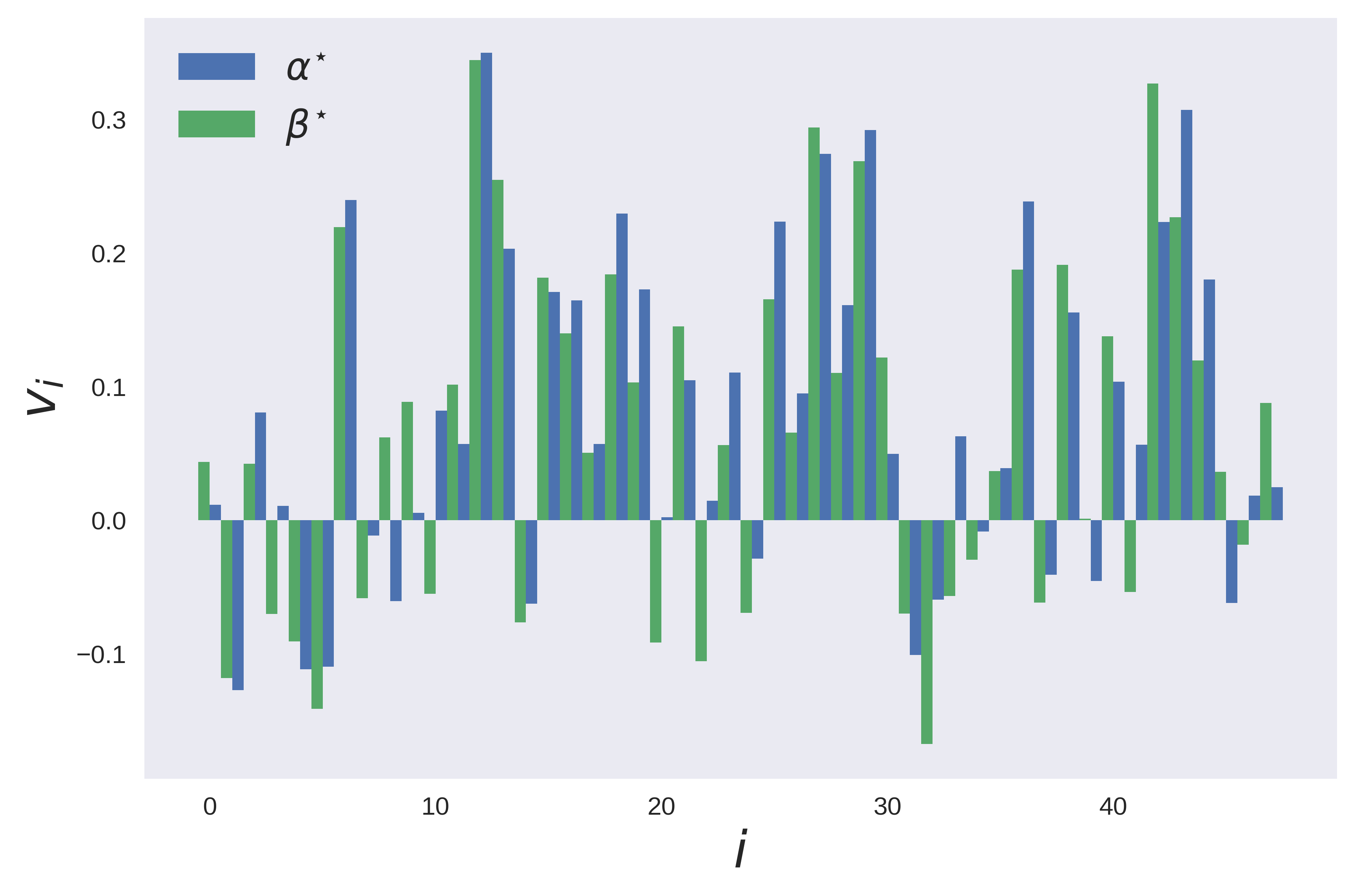}
    \includegraphics[width=8cm]{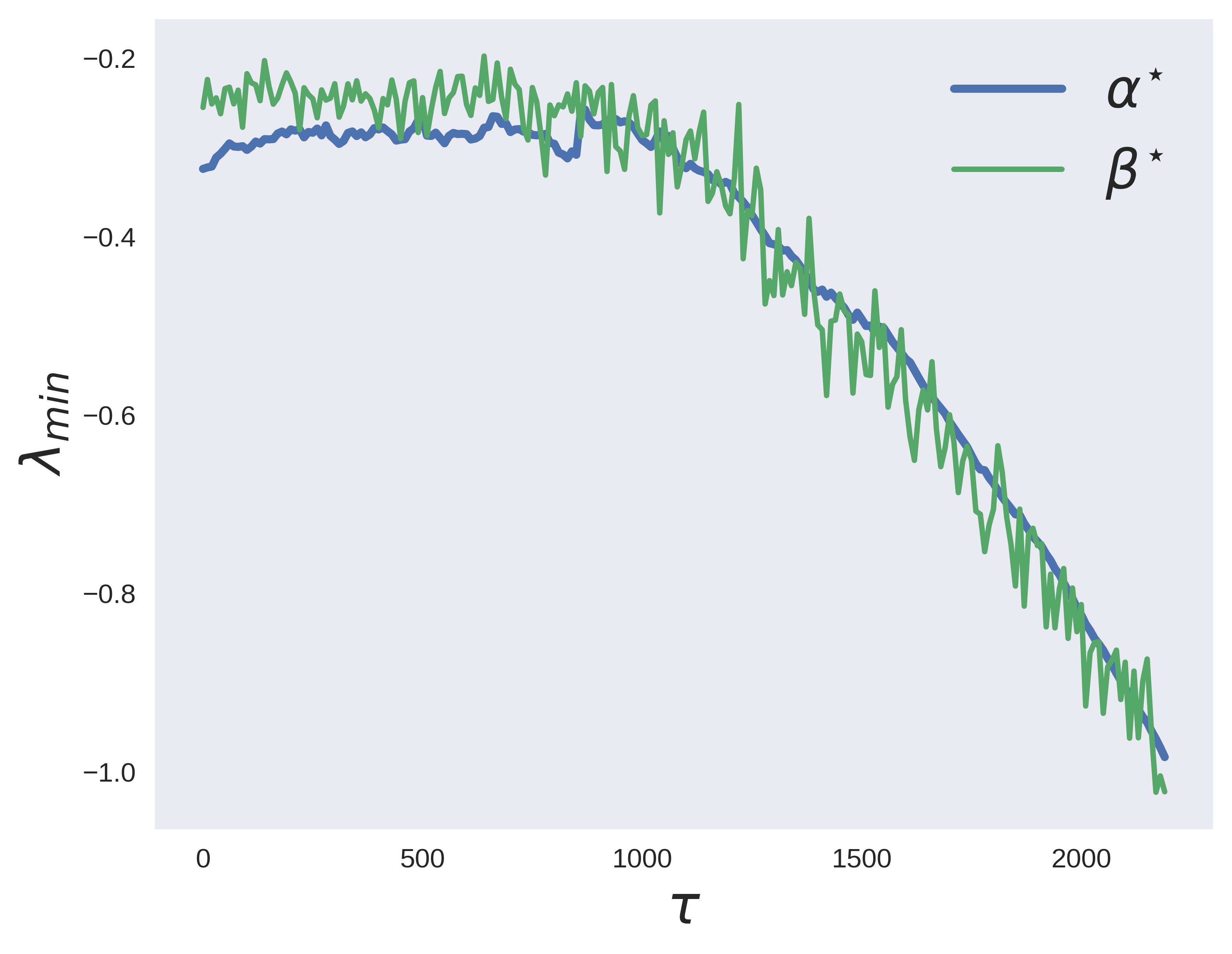}
    \caption{\textbf{Evolved and synthetic matrix have similar structure of eigenstates.} Comparison between the eigenvalues of the evolved interaction matrix $\alpha^{\star}$ (blue) and of the maximum entropy matrix $\beta^{\star}$ (green) at generation $\tau=1500$ (Top) and the coefficient of the eigenvector of the minimal eigenvalue (middle). Evolution of the minimum eigenvalue of both matrices for every generations (Bottom).}
    \label{SI:fig:max_ent}
\end{figure}

Following \cite{barbierFingerprintsHighDimensionalCoexistence2021}, we can compute the inferred interaction matrix $\beta$ that gives rise to the observed equilibrium abundances and has a given mean interaction $\bar{\beta}$. Neglecting the correlations between different matrix elements (which are sub-leading in the large $S$ limit), such matrix is of the form:
\begin{equation}
    \beta^{\star}_{ij}=\bar{\beta}+\frac{(K_i-N_i-\bar{\beta})N_j}{\sum_kN_k^2}+\sigma B_{ij}
\end{equation}
with $B$ a Gaussian matrix of mean $0$ and variance $1$. Fig. \ref{SI:fig:max_ent} shows the eigenvalue distribution of such matrix, compared to the one obtained from the evolutionary process. The agreement is remarkable. 

\section{Generalisations}\label{SI:generalizations}

We checked that the features presented in the main text hold for a broad range of parameters in the numerical simulations, in particular for most initial values of $(\mu, \sigma)$ as long as we are in the unique equilibrium phase (see section \ref{SI:section:phase}), for any values of $m$ and $n$ (when selecting $m$ communities out of $n$) as long as $n>1$ and $m<n$ (theses two extreme cases lead to no selection) and for any value of $\gamma$ with the exception of $\gamma=-1$, as expected from the equations discussed in the main text. The addition of a small immigration term to the ecological dynamics, moreover, doesn't qualitatively alter the results.

We considered different selection functions, such as $f=\sum_iw_iN_i$ with all weights $w_i$ being positives. Similar results keep holding. This can be understood by interpreting the selection process as a modification of selection for the total abundance, obtained by rescaling species abundances as $w_iN_i$, so that species $i$ has carrying capacities $w_iK_i$ and an interaction matrix $\alpha_{ij}w_i/w_j$. 
The same computations explained in the main text and in Section \ref{SI:section:mut_sel} then allow us to obtain the exact same recursive equations for $f$ and $\alpha$ but with $\vb{v}=\frac{\delta f}{\delta K}$ and $\vb{v}^{\star}(\tau)=(\mathbb{I}^{\star}+\alpha^{\star}(\tau)^{\top})^{-1}\vb{w}^{\star}$.

\section{Numerical integration method} \label{SI:numerical_considerations}
We here present an integration method for the Lotka-Volterra equations :
\begin{equation}\label{SI:LV_eq}
\frac{\mathrm{d}N_i}{\mathrm{d}t} = r_i\frac{N_i}{K_i}\left(K_i-N_i-\sum_{j\neq i}\alpha_{ij}N_j\right),
\end{equation}
with $K_i$ the carrying capacities, $r_i$ the growth rates (all equals to one in the paper) and $\alpha$ the interaction matrix.

In contrast to the Euler method where one assumes that the derivative is constant during a short interval of time $\mathrm{d}t$, potentially leading to negative populations for strong derivatives, we assume that only the abundances of the other species $(N_j(t))_{j\neq i}$ are constant. During this time interval $[t, t+\mathrm{d}t]$, the equations are reduced to uncoupled logistic equations of effective carrying capacities $\overset{\sim}{K_i}= K_i-\sum_{j\neq i}\alpha_{ij}N_j(t)$ and effective growth rates $\overset{\sim}{r_i}(t)=r_i\overset{\sim}{K_i}(t)/K_i$. These logistic equations can be analytically solved in the interval $[t, t+\mathrm{d}t]$, giving the recursive formula:
\begin{equation}
N_i(t+\mathrm{d}t)=\frac{N_i(t)\overset{\sim}{K_i}(t)}{N_i(t)+(\overset{\sim}{K_i}(t)-N_i(t))\exp(-\overset{\sim}{r_i}(t)\mathrm{d}t)}
\end{equation}
with \begin{equation}
\begin{cases}
\overset{\sim}{K_i}(t) = K_i-\sum_{j\neq i}\alpha_{ij}N_j(t)\\
\overset{\sim}{r_i}(t)=r_i\frac{\overset{\sim}{K_i}(t)}{K_i}
\end{cases}
\end{equation}

We then have a logistic-by-part curve that avoids abundances to become negative. We can show by Taylor expansion that we get back the Euler method at first order in $\mathrm{d}t$.

\bibliography{biblio.bib}
\bibliographystyle{apalike}